\documentclass[]{jfm}

\usepackage{natbib}

\ifx\pdfoutput\undefined
\usepackage{graphicx}
\DeclareGraphicsExtensions{.ps,.eps,.cps}
\newcommand\DOT{.}
\else
\usepackage[pdftex]{graphicx}
\DeclareGraphicsExtensions{.png,.pdf}
\newcommand\DOT{.}
\fi

\ifCUPmtlplainloaded \else
  \checkfont{eurm10}
  \iffontfound
    \IfFileExists{upmath.sty}
      {\typeout{^^JFound AMS Euler Roman fonts on the system,
                   using the 'upmath' package.^^J}%
       \usepackage{upmath}}
      {\typeout{^^JFound AMS Euler Roman fonts on the system, but you
                   dont seem to have the}%
       \typeout{'upmath' package installed. JFM.cls can take advantage
                 of these fonts,^^Jif you use 'upmath' package.^^J}%
       \providecommand\upi{\pi}%
      }
  \else
    \providecommand\upi{\pi}%
  \fi
\fi

\ifCUPmtlplainloaded \else
  \checkfont{msam10}
  \iffontfound
    \IfFileExists{amssymb.sty}
      {\typeout{^^JFound AMS Symbol fonts on the system, using the
                'amssymb' package.^^J}%
       \usepackage{amssymb}%

      }{}
  \fi
\fi

\ifCUPmtlplainloaded \else
  \IfFileExists{amsbsy.sty}
    {\typeout{^^JFound the 'amsbsy' package on the system, using it.^^J}%
     \usepackage{amsbsy}}
    {\providecommand\boldsymbol[1]{\mbox{\boldmath $##1$}}}
\fi

\providecommand\bnabla{\boldsymbol{\nabla}}
\providecommand\bcdot{\boldsymbol{\cdot}}

\newcommand\Rey{\mbox{\textit{Re}}}  

\newcommand\slsA{\mathsfbi{A}} 
\newcommand\slsB{\mathsfbi{B}} 
\newcommand\slsI{\mathsfbi{I}} 
\newcommand\Diff{\mathsfbi{D}} 

\newcommand\bita{\mathit{\boldsymbol{a}}}   
\newcommand\bitb{\mathit{\boldsymbol{b}}}   
\newcommand\bitC{\mathit{\boldsymbol{C}}}   
\newcommand\bitQ{\mathit{\boldsymbol{Q}}}   
\newcommand\bitF{\mathit{\boldsymbol{F}}}   
\newcommand\bitG{\mathit{\boldsymbol{G}}}   
\newcommand\bitu{\mathit{\boldsymbol{u}}}   
\newcommand\bitv{\mathit{\boldsymbol{v}}}   
\newcommand\bitx{\mathit{\boldsymbol{x}}}   
\newcommand\bitr{\mathit{\boldsymbol{r}}}   
\newcommand\bitt{\boldsymbol{\theta}}   
\newcommand\bitphi{\boldsymbol{\phi}}   
\newcommand\bitz{\mathit{\boldsymbol{z}}}   

\newcommand\sbA{\mathcal{{A}}}   
\newcommand\sbM{\mathcal{{M}}}   
\newcommand\sbx{\mathrm{{x}}}   

\newcommand\sLambda{{\it \Lambda}}   

\newsavebox{\astrutbox}
\sbox{\astrutbox}{\rule[-5pt]{0pt}{20pt}}

\newcommand\eg{e.g.\ }
\newcommand\ie{i.e.\ }

\graphicspath{{./}{./Figs/}}

\title[Takens-Bogdanov of travelling waves]{Takens-Bogdanov bifurcation of
  travelling wave solutions in pipe flow}

\author[F. Mellibovsky and B. Eckhardt]%
{F.\ns M\ls E\ls L\ls L\ls I\ls B\ls O\ls V\ls S\ls K\ls Y$^1$
\and B.\ns E\ls C\ls K\ls H\ls A\ls R\ls D\ls T$^2$}

\affiliation{$^1$Departament de F{\'{i}}sica Aplicada, Universitat
  Polit{\`{e}}cnica de Catalunya, 08034, Barcelona, Spain
  \\[\affilskip] $^2$Fachbereich Physik, Philipps-Universit\"{a}t
  Marburg, D-35032 Marburg, Germany}

\pubyear{????}
\volume{???}
\pagerange{???--???}
\date{?? and in revised form ??}

\begin{document}

\maketitle

\begin{abstract}
The appearance of travelling-wave-type solutions in pipe Poiseuille
flow that are disconnected from the basic parabolic profile is
numerically studied in detail.  We focus on solutions in the $2$-fold
azimuthally-periodic subspace because of their special stability
properties, but relate our findings to other solutions as well.  Using
time-stepping, an adapted {\em Krylov}-Newton method and Arnoldi
iteration for the computation and stability analysis of relative
equilibria, and a robust pseudo-arclength continuation scheme we
unfold a double-zero (Takens-Bogdanov) bifurcating scenario as a
function of Reynolds number ($\Rey$) and wavenumber ($\kappa$). This
scenario is extended, by the inclusion of higher order terms in the
normal form, to account for the appearance of supercritical modulated
waves emanating from the upper branch of solutions at a degenerate
Hopf bifurcation. These waves are expected to disappear in saddle-loop
bifurcations upon collision with lower-branch solutions, thereby
leaving stable upper-branch solutions whose subsequent secondary
bifurcations could contribute to the formation of the phase space
structures that are required for turbulent dynamics at higher $\Rey$.
\end{abstract}

\section{Introduction}\label{sec:intro}

Subcritical transition and the sustainment of turbulence at moderate
flow rates in shear flows is a problem of great theoretical complexity
and practical relevance \cite[][]{Gro00,ESHW91,Eckhardt09}.  Since
Osbourne \cite{Rey1883} much celebrated work on the onset of turbulent
motion in a straight pipe of circular cross-section, this problem has
become a benchmark of transition without linear instability, \ie
subcritical transition. Many theoretical \cite[][]{BoBr88,BrGr99},
numerical \cite[][]{ScHe_JFM_94,Zik_PoF_96,SMZN_JFM_99} and
experimental
\cite[][]{WyCh_JFM_73,WySoFr_JFM_75,DaMu_JFM_95,HSWE_NATURE_06}
efforts have been devoted at comprehending how the laminar state
(Hagen-Poiseuille flow) becomes unstable to finite amplitude
perturbations despite its linear stability
\cite[][]{Pfenniger_B_61,MeTr_JCP_03}.

The numerical computation of finite amplitude secondary solutions in
the form of travelling waves \cite[][]{FaEc_PRL_03,WeKe_JFM_04}, whose
existence relies on a mechanism called the self-sustained process
\cite[][]{Wal_PoF_97}, initially advanced to account for turbulent
regeneration, has brought renewed interest to the problem within the
dynamical systems community. These waves typically appear in
saddle-node bifurcations and break some of the symmetries of the
problem while retaining others
\cite*[][]{PrKe_PRL_07,PrDuKe_PTRSA_09}. Their lower branches have
been repeatedly shown to dwell on the critical threshold
\cite*[][]{DuWiKe_JFM_08,MeMe_PTRSA_09} separating the basins of
attraction of
laminar and turbulent flows, which suggests their implication in the
transition process. Upper-branch solutions, with wall frictions
closer to turbulence, seem to play a role in developed turbulence
\cite*[][]{ScEcVo_PRE_07,KeTu_JFM_07} and have been observed in experiments
\cite[][]{HVWNFEWKW_SCI_04}.

The $2$-fold discrete azimuthally-periodic subspace constitutes a
special case in that it contains a family of travelling waves with
unusually low-dimensional unstable manifolds.
All other known families of travelling waves exhibit at least one
unstable mode within their subspace that is shared by all waves,
thereby rendering the family globally unstable.  This is not the case
of a specific shift-reflect, $2$-fold azimuthally-periodic family of
travelling waves, the special stability properties of which can be
qualified as {\em convenient}.  Lower-branch solutions of this family
possess a single unstable direction within the subspace, which
ultimately assists bare symmetry-restricted time evolution in
converging them numerically \cite*[][]{DuWiKe_JFM_08}, when adequately
combined with edge tracking techniques
\cite*[][]{SkYoEc_PRL_07,ScEcYo_PRL_07}. The lower branch being a
saddle, it would not be surprising if the upper branch constituted a
stable node, at least for some of the parameter values. Additional
unstable directions arise when analysing the stability in full space
or when considering subharmonic instability to longer wavelengths, but
the convenient stability properties within their azimuthal subspace,
which can by itself sustain turbulence, renders it a particularly nice
playground for the search of all sorts of time-dependent solutions
with simple numerical machinery. These solutions may help understand
turbulence within the subspace and give hints as to what might be
happening in full space.


This study aims at analysing and describing the phenomena associated
to the emergence of disconnected solutions in shear flows. While the
current analysis focuses on a given family of travelling waves in
circular pipe flow, the results are thought to be common feature in a
vast range of other travelling wave families in pipe flow and several
equilibria and relative equilibria in a variety of other shear flows
such as plane Couette \cite*[][]{Nagata_PRE_97,WaGiWa_PRL_07}
or plane Poiseuille \cite*[][]{PuSa_JFM_88,SoMe_JFM_91,EhKo_JFM_91}
flows.


The outline of the paper is as follows. In \S\ref{sec:formet}, we
present the pipe Poiseuille flow problem and sketch a numerical scheme
for the integration of the resulting equations. The symmetries of the
problem are discussed and numerical methods for the computation and
stability analysis of relative equilibria are introduced. Some aspects
of normal form reduction in the case of bifurcating relative
equilibria are also reviewed in \S\ref{sec:formet}. The main results
of the computations are presented in \S\ref{sec:results}, where the
different types of solutions resulting from an extensive parameter
space exploration are exhibited, and evidence of a double-zero
bifurcation scenario is exposed. These results are thoroughly
discussed and explained in \S\ref{sec:discu} by considering the
inclusion of higher order terms in the relevant normal
form. Conclusions and ensuing prospects and challenges are summarised
in \S\ref{sec:conclu}.

\section{Formulation and methods}\label{sec:formet}

\subsection{Equations and numerical scheme}\label{sec:eqnumsch}


The constant flow-rate motion of an incompressible Newtonian fluid
through a straight pipe of circular cross-section is considered. The
governing parameter is the Reynolds number, defined as $\Rey = U D /
\nu$, where $U$ is the mean axial flow speed, $D$ the pipe diameter
and $\nu$ the kinematic viscosity of the fluid. The dynamics is
governed by the incompressible Navier-Stokes equations.

In cylindrical nondimensional coordinates ${\bitx}=(r,\theta,z)$, the
basic Hagen-Poiseuille solution
 of the Navier-Stokes problem reads ${\bitu}_{\rm b}=u_r^{\rm
b}\;\hat{\bitr}+u_{\theta}^{\rm b}\;\hat{\bitt}+u_z^{\rm
b}\;\hat{\bitz}=(1-r^2)\;\hat{\bitz}$. The Navier-Stokes equations for
the velocity-pressure ${\bitu}=(u_r,u_{\theta},u_z)$-$p$ perturbation
fields are
\begin{equation}
  \partial_t{\bitu} = -\bnabla p + \frac{1}{\Rey}\Delta {\bitu} -
  ({\bitu} \bcdot \bnabla) ({\bitu}_{\rm b} +{\bitu}) - ({\bitu}_{\rm b}
  \bcdot \bnabla) {\bitu} + f \; \hat{\bitz},\\
  \label{eq:NS}
\end{equation}
\begin{equation}
  \bnabla \bcdot  {\bitu} = 0,
  \label{eq:DV}
\end{equation}
\begin{equation}
  Q(\bitu) = \int_0^{2 \pi} \int_0^1 (\bitu \bcdot \hat{\bitz}) \; r
  {\rm d}r {\rm d}\theta = 0,
  \label{eq:CMc}
\end{equation}
\begin{equation}
  {\bitu}(1,\theta,z;t)={\boldsymbol 0}, \;\; {\bitu}(r,\theta+2\pi/{n_s},z,t)={\bitu}(r,\theta,z+2\upi/\kappa,t)={\bitu}(r,\theta,z,t)
\label{eq:BC},
\end{equation}
where an adjustable axial forcing $f=f(t)$ in (\ref{eq:NS}) ensures
the constant mass-flux constraint (\ref{eq:CMc}). The non-slip
boundary condition at the wall and the axial and azimuthal
periodicities are enforced by (\ref{eq:BC}), where $\kappa$ is the
non-dimensional wavenumber, here considered as an additional parameter
selecting the axial periodicity of the flow structures. In this study,
the axial wavenumber is varied in the range $\kappa \in [1.1,1.85]$,
corresponding to pipe periodicities $\sLambda \in [3.40,5.71]$, in
units of $D/2$.
This range covers all relevant existing solutions up
to $\Rey \simeq 1500$ while discarding subharmonic instabilities that
might complicate the bifurcation scenario under examination at higher $\Rey$.
The azimuthal wavenumber is set to $n_s=2$ throughout, meaning that
only $2$-fold azimuthally-periodic fields are considered. All
solutions living in this subspace are also solutions to the full
Navier-Stokes equations with $n_s=1$, but the instabilities associated
to perturbations without the $n_s=2$ symmetry are suppressed,
rendering some of the solutions stable and thus accessible through
time evolution.

For the spatial discretisation of (\ref{eq:NS}-\ref{eq:BC}) we use a
solenoidal spectral Petrov-Galerkin scheme thoroughly described and
tested by \cite*{MeMe_ANM_07}. The method implicitly fulfils
divergence free and boundary conditions, and dispenses with the need
for dealing with the pressure field. The velocity field is expanded
following
\begin{equation}
  \bitu(r,\theta,z;t)=\sum_{l=-L}^{L} \sum_{n=-N}^{N}
  {e}^{-{\rm i} (\kappa l z + n_s n \theta)} \;\bitu_{ln}(r;t)
\label{eq:uexp1},
\end{equation}
\begin{equation}
  \bitu_{ln}(r;t)=
  \sum_{m=0}^{M} a_{lnm}^{(1,2)}(t) \; \bitv_{lnm}^{(1,2)}(r)
\label{eq:uexp2},
\end{equation}
with $a_{lnm}^{(1,2)}$ the complex expansion coefficients defining the
vector state $\bita$. In this way, the field decouples into its
axial-azimuthal Fourier components $\bitu_{ln}$. As already noted, the
expansion basis $\bitv_{lnm}(r)$, presented in the work of
\citet[eq. (24-27)]{MeMe_ANM_07}, is chosen so that boundary
conditions (\ref{eq:BC}) and solenoidality (\ref{eq:DV}) are
implicitly satisfied. The spectral resolution, checked as adequate for
the computations performed in this study, has been set to $L=16$,
$N=12$ and $M=36$, corresponding to $\pm 16$ axial and $\pm 24$
azimuthal Fourier modes, and to $37$ Chebyshev collocation points for
the radial coordinate. Upon projection on the test basis
$\tilde{\bitv}_{lnm}(r)$ \cite[][eq. (32-35)]{MeMe_ANM_07}, we are left
with a system of ordinary differential equations for the
complex-valued state vector $\bita$:
\begin{equation}
  \slsA \; \dot{\bita} = \left( \frac{1}{\Rey} {\slsB}_1 +
  {\slsB}_2 \right) \; \bita - \bitb(\bita,\bita) + f \; \bitC,
\label{eq:ODEs}
\end{equation}
\begin{equation}
  {\bitQ} \bcdot \bita = 0,
\label{eq:CM}
\end{equation}
where the subscripts in \citet[eq. (47)]{MeMe_ANM_07} have been
omitted for simplicity and $\slsB$ has been decomposed into two
separate contributions, ${\slsB}_1$ corresponding to the viscous
dissipation term, with the dependence on $\Rey$ rendered explicit, and
${\slsB}_2$ to the linear advection terms. $\bitC$ is the projection
of $\hat{\bitz}$ onto the test basis and is therefore responsible for
keeping the mass-flux constant, represented by the linear scalar
equation (\ref{eq:CM}), by instantaneous adjustment of $f$.

The system (\ref{eq:ODEs}) is evolved in time using a 4th order linearly
implicit method with $\Delta t = 1 \times 10^{-2}$, where time is
measured in units of $D/(4 U)$.

It will be handy to define some global properties of the flow that are
independent of solid-body motions such as translation or rotation. One
such quantity is the normalised energy of any given perturbation field
${\bitu}$, which is given by
\begin{equation}
    \varepsilon({\bitu}) = \frac{1}{2\epsilon_{\rm b}} \int_0^{2
    \upi/\kappa} {\rm d}z \int_0^{2 \pi} {\rm d}\theta \int_0^1 r{\rm
    d}r \; {\bitu}^* \bcdot {\bitu}
\label{eq:energy1},
\end{equation}
with $\epsilon_{\rm b}=\upi^2/(3\kappa)$ the energy of the basic flow
and $^*$ symbolising complex conjugation. This energy decouples
exactly into the sum of its axial-azimuthal Fourier components:
\begin{equation}
    \varepsilon({\bitu}) = \sum_{l=-L}^{L} \sum_{n=-N}^{N}
	\varepsilon_{ln} = \varepsilon_{\rm 1D} + \varepsilon_{\rm 2t}
	+ \varepsilon_{\rm 2z}+ \varepsilon_{\rm 3D}
\label{eq:energy2},
\end{equation}
where $\varepsilon_{ln}=\varepsilon(\bitu_{ln})$ is the energy
associated to the $(l,n)$ axial-azimuthal Fourier mode. To the right
of (\ref{eq:energy2}) the Fourier modes have been grouped in
meaningful sets corresponding to the axisymmetric
streamwise-independent component ($\varepsilon_{\rm 1D}$, mean flow
component), the non-axisymmetric streamwise-independent component
($\varepsilon_{\rm 2z}$, energy signature of vortices and streaks),
the axisymmetric streamwise-dependent component ($\varepsilon_{\rm
2\theta}$, mean wavy component) and the non-axisymmetric
streamwise-dependent component ($\varepsilon_{\rm 3D}$, purely
three-dimensional energy).

Another global property is the mean axial pressure gradient that is
required to drive the flow at constant mass-flux, relative to its
laminar flow value. It can be easily stated in terms of the adjusting
intensity of the axial forcing $f$ and of $\Rey$:
\begin{equation}
  (\bnabla p)_z = 1+\frac{Re f}{4}
\label{eq:gradp}.
\end{equation}

Finally, it will be useful to define the instantaneous axial and
azimuthal phase speeds for any given velocity field. Their definition
is clear for solid-body moving solutions, but can be extended to any
velocity field by decomposing the time-evolving state vector into two
contributions: one associated to solid-body motion and the other to a
modulation. This decomposition can be made unique by minimising, as a
function of the phase speeds, at any given time, the $2$-norm of the
modulational component:
\begin{equation}
  \min_{c_z,c_{\theta}} \left\Vert a(t+\Delta t) - a(t) \; \mathrm{e}^{-{\rm
  i}(\kappa l c_z + n_s n c_{\theta}) \Delta t} \right\Vert
  \label{eq:PhSpeeds}
\end{equation}
For all rotating and travelling waves, $c_z$ and $c_{\theta}$ are
constant, the minimum (\ref{eq:PhSpeeds}) is exactly zero, and they
coincide with the axial and azimuthal phase speeds, respectively.  The
effective computation is done instantaneously by considering two
consecutive time instants separated by $\Delta t$ and by applying a
root-finding Newton method to the $2$-dimensional Jacobian of the norm
in (\ref{eq:PhSpeeds}).

\subsection{Symmetries of the problem}

Pipe flow and its basic solution possess $O(2) \times SO(2)$ symmetry,
\ie it is invariant under all azimuthal rotations about the axis and
reflections with respect to all diametral planes, as well as under all
axial translations. All travelling wave families discovered so far are
periodic and have axial and azimuthal dependency, thus breaking all
continuous symmetries. While, in the azimuthal direction, the natural
periodicity of the pipe renders all solutions invariant under
rotations of $2 \upi$, travelling waves can have an additional
discrete symmetry that renders them invariant under the cyclic group
$Z_{n_s}=\{R_{2\upi},R_{2\upi/2},...,R_{2\upi/n_s}\}$ with $n_s$ the
azimuthal wave number and
\begin{equation}
R_{\alpha}({\bitu})({\bitx})=R_{\alpha}(u,v,w)(r,\theta,z;t)=(u,v,w)(r,\theta+\alpha,z;t)
\label{eq:RotSym}.
\end{equation}
In the axial direction, on top of the trivial shift by a wavelength,
the axial symmetry rupture gives rise to a special type of time
periodicity, in the sense that the time dependence is a bare drift in
the direction of the $SO(2)$ symmetry. Travelling waves are therefore
best described as relative equilibria that posses the continuous
space-time symmetry
\begin{equation}
(u,v,w)_{\rm tw}(r,\theta,z;t)=(u,v,w)_{\rm tw}(r,\theta,z - c_z t;0)
\label{eq:SpaceTimeSym},
\end{equation}
where $c_z$ is the axial drift speed. In a comoving reference frame
travelling downstream with speed $c_z$, travelling waves appear as
stationary solutions. Near a relative equilibrium the drift dynamics
is trivial and decouples from the dynamics orthogonal to the drift. As
a result, bifurcations of relative equilibria can be analysed in two
steps, first describing the bifurcations associated to the orthogonal
dynamics, then adding the corresponding drift along the travelling
direction \cite*[][]{Krupa_SJMA_90}.

There are a couple of additional non-trivial symmetries that leave
some of the known travelling waves unaltered. The first one is a
mirror symmetry with respect to $n_s$ diametral planes:
\begin{equation}
M{\bitu}({\bitx})=M(u,v,w)(r,\theta_i+\theta,z;t)=(u,-v,w)(r,\theta_i-\theta,z;t)
\label{eq:MirrorSym},
\end{equation}
with $\theta_i=\theta_0+i\upi/n_s$, $i=\{0,1,...,n_s-1\}$, and
$\theta_0$ parametrising the azimuthal degeneracy of solutions.

The second one, characteristic of the travelling wave family studied
here, is a combined shift-reflect symmetry. Solutions invariant under
this symmetry operation,
\begin{equation}
S{\bitu}({\bitx})=S(u,v,w)(r,\theta_i+\theta,z;t)=(u,-v,w)(r,\theta_i-\theta,z+\upi/\kappa;t)
\label{eq:ShiftRefSym},
\end{equation}
are left unaltered when shifted half a wavelength downstream and then
reflected with respect to any of $n_s$ diametral planes tilted with
$\theta_i$.

A number of highly symmetric solutions possessing both symmetries
\cite[][]{PrDuKe_PTRSA_09} have been identified and continued to very
low $\Rey$ numbers. Some of the less symmetric waves are known to
emerge from symmetry breaking bifurcations of these highly symmetric
waves.

All symmetries can be enforced upon computation. This obviously
applies to time evolution, computation of relative equilibria using
Newton iteration and stability analysis of relative equilibria via
Arnoldi iteration. Nevertheless, we have chosen to only restrict the
azimuthal wave number $n_s$, apart from the invariance due to axial
periodicity, to avoid inevitable instability to azimuthally
non-symmetric flows. Every additional symmetry has been exposed to
eventual break-up.

\subsection{Computation of relative equilibria, continuation and stability analysis}

The spectral method discussed in section \ref{sec:eqnumsch} is
extremely versatile and very well suited for straightforward coding of
a Newton method for the computation of relative equilibria, for their
continuation in parameter space and for linear stability analysis
using Arnoldi iteration. Some of these usages of the spectral method
were briefly described in \cite{MAMM_EPJST_07} but will be
developed further here.

\subsubsection{Computation of travelling/rotating waves}
The spectral representation of a generic travelling and/or rotating
wave reduces to
\begin{equation}
  \bita(t) = \bita_{\rm tw} \mathrm{e}^{-{\rm i}(\kappa l c_z + n_s n
  c_{\theta}) t},
  \label{eq:TrWav}
\end{equation}
where $\bita_{\rm tw}$ is the state vector corresponding to the
time-independent structure of the wave and $c_z$ and $c_{\theta}$ are
the axial and azimuthal advection speeds, respectively. The
travelling-waves family studied here has no rotation and, therefore,
$c_{\theta}=0$.

Substitution of (\ref{eq:TrWav}) into (\ref{eq:ODEs},\ref{eq:CM})
yields a non-linear system of algebraic equations
\begin{equation}
  \left\{
  \begin{array}{l}
    \bitF(\bita_{\rm tw},c_z,c_{\theta},f) = {\bf 0}\\
	 \bitQ \bcdot \bita_{\rm tw} = 0 \\
	 \phi_z(\bita_{\rm tw}) = \phi_z(\bita^0) \\
	 \phi_{\theta}(\bita_{\rm tw}) = \phi_{\theta}(\bita^0)
  \end{array}
  \right.
  \label{eq:Ftw},
\end{equation}
where the last two scalar equations impose the phase on any two
suitable coefficients so as to lift the translational and rotational
degeneracy of the solution. The vector function $\bitF$ is given by
\begin{equation}
  \bitF(\bita_{\rm tw},c_z,c_{\theta},f)=\left( \frac{1}{\Rey} {\slsB}_1 +
  {\slsB}_2 + {\rm i}(\kappa l c_z + n_s n c_{\theta}) {\slsA}
  \right) \; \bita_{\rm tw} - \bitb(\bita_{\rm tw},\bita_{\rm tw}) + f \; \bitC
  \label{eq:Funct}.
\end{equation}

System (\ref{eq:Ftw}) can be solved via Newton iteration. At step $k$,
the linear system to be solved takes the form
\begin{equation}
  \left[
    \begin{array}{cccc}
      {\Diff}_\bita \bitG^k &
      {\Diff}_{c_z} \bitG^k &
      {\Diff}_{c_{\theta}} \bitG^k &
      \Rey {\slsB}_1^{-1} \bitC \\
      \bitphi_z^{T} &
      0 &
      0 &
      0 \\
      \bitphi_{\theta}^{T} &
      0 &
      0 &
      0 \\
      {\bitQ}^{T} &
      0 &
      0 &
      0
    \end{array}
    \right]
  \left(
    \begin{array}{l}
      \Delta \bita^k \\
      \Delta c_z^k \\
      \Delta c_{\theta}^k \\
      \Delta f^k
    \end{array}
    \right) =
  - \left(
    \begin{array}{c}
      \bitG^k \\
      0 \\
      0 \\
      0
    \end{array}
    \right)
  \label{eq:Newton},
\end{equation}
where Stokes preconditioning (multiplication by $\Rey {\slsB}_1^{-1}$
from the left) has been applied to the first row to assist convergence
\cite*[][]{MaTu_PoF_95}.
 Accordingly, $\bitG^k=\Rey {\slsB}_1^{-1}
\bitF(\bita^k,c_z^k,c_{\theta}^k,f^k)$. The terms appearing on the
first row of the left hand side correspond to the Jacobian of $\bitG$,
noted $\Diff_x \bitG$, evaluated at step $k$:
\begin{equation}
  \left\{
  \begin{array}{l}
      {\Diff}_\bita \bitG^k =  {\slsI} + \Rey {\slsB}_1^{-1} \slsB_2 + {\rm i} \Rey (\kappa l c_z^k + n_s n c_{\theta}^k){\slsB}_1^{-1}{\slsA} - \Rey {\slsB}_1^{-1} {\Diff}_\bita \bitb(\bita^k)\\
      {\Diff}_{c_z} \bitG^k =  {\rm i} \kappa l \Rey {\slsB}_1^{-1}{\slsA} \bita^k \\
      {\Diff}_{c_{\theta}} \bitG^k =  {\rm i} n_s n \Rey {\slsB}_1^{-1}{\slsA} \bita^k
  \end{array}
  \right.
  \label{eq:Jacobian},
\end{equation}
where ${\Diff}_\bita \bitb(a^k) \Delta \bita^k$, the linearised
nonlinear term, is computed in physical space as $(\Delta {\bitu}^k
\bcdot \bnabla){\bitu_{\rm tw}}^k + ({\bitu_{\rm tw}}^k \bcdot
\bnabla)\Delta {\bitu}^k$ and then projected back to spectral space
\cite[][]{MeMe_ANM_07}. Arrays $\bitphi_z$ and $\bitphi_{\theta}$
correspond to the linearisation of the phase prescription of any two
coefficients, as long as they are associated with basis elements that
entrain axial and azimuthal structure, respectively. In the notation
of \cite{MeMe_ANM_07}, we have chosen to fix the phase of
$a_{100}^{(2)}$ and $a_{010}^{(2)}$ to lift the axial and azimuthal
degeneracies of the solutions, respectively. The incremental vector
the Newton method requires solving for at each step is defined as
$\Delta x^k = x^{k+1} -x^k$ and provides the new guess for step $k+1$.

The solution of the linear system at every Newton step needs only be
approximated by using inexact Krylov iteration. Generalised minimal
residuals \cite*[see][for the implementation of GMRES we
  use]{CERFACS_03} and stabilised biconjugate gradient methods
(BiCGSTab) can be easily implemented \cite*[see \eg][and references
  therein]{QuSaSa_B_07}. Convergence to the exact solution of the
nonlinear equations to numerical accuracy is guaranteed provided that
a sufficiently good initial guess $(\bita^0,c_z^0,c_{\theta}^0,f^0)$
is fed into the iteration. Convergence can be enhanced by appropriate
damping \cite[][]{QuSaSa_B_07} or trust-region techniques
\cite*[][]{DeSc_B_96}.

\subsubsection{Continuation of solutions}
It is fairly easy to extend the Newton method just described to the
continuation in $\Rey$ of any branch of solutions. The explicit
dependence of system (\ref{eq:Ftw}) on $\Rey$ allows this parameter to
be treated as an additional unknown. Orthogonality to the Jacobian at
some prescribed distance (pseudo-arclength) from the departing
solution provides the additional equation needed
\cite*[][]{SaMaLo_JCP_02}. This continuation technique is capable of
following branches around turning points, which makes it very
convenient in the presence of saddle-node bifurcations. Continuation
in the other parameter $\kappa$ is not as straight forward. The
complex dependency of the linear contribution of $\bitF$ on $\kappa$
requires recomputing the linear operators at every single Krylov
iteration, which renders the method very costly. We have chosen to
continue it manually by simply using the Newton method with discrete
jumps in $\kappa$, small enough so that the converged solution at the
previous $\kappa$ constitutes a sufficiently good guess. This
procedure cannot go around turning points, but combined with
pseudo-arclength continuation in $\Rey$, the full family of travelling
waves can be computed.

\subsubsection{Stability analysis}
Special care must be taken when analysing the linear stability of
rotating/travelling waves. Waves of this type have two equivalent
representations in the constant pressure gradient and in the constant
flow-rate frames. This is so because their solid-body motion neither
modifies the mass-flux nor the mean axial pressure gradient. However,
their destabilisation does not necessarily preserve these two global
quantities simultaneously, meaning that linear stability will
explicitly depend on which of the two is prescribed as constant. Since
we have chosen to study pipe flow at a constant flow rate, we must
accordingly allow for pressure variation. Let us consider an
infinitesimal perturbation ($\bitv_\bita$,$v_f$) on a given
travelling/rotating wave ($\bita_{\rm tw}$,$f_{\rm tw}$) in a
reference frame moving with the wave:
\begin{equation}
  \left\{
  \begin{array}{l}
    \bita = \left( \bita_{\rm tw}+\bitv_\bita \mathrm{e}^{\lambda t}
    \right) \mathrm{e}^{-{\rm i}(\kappa l c_z + n_s n c_{\theta}) t} \\
    f = f_{\rm tw} + v_f \mathrm{e}^{\lambda t}
  \end{array}
  \right.
  \label{eq:infpert}.
\end{equation}
The factor $\mathrm{e}^{\lambda t}$ is intended to capture the linear
evolution of the perturbation in the vicinity of the relative
equilibrium. In this sense, ($\bitv_\bita$,$v_f$) is an eigenvector of
eigenvalue $\lambda$, which in general can be complex. Introducing
(\ref{eq:infpert}) into (\ref{eq:ODEs},\ref{eq:CM}) yields the
evolution equation for the perturbation, which can be linearised
around the wave solution by discarding all second order terms:
\begin{equation}
  \left\{
  \begin{array}{rcl}
    \left( \frac{1}{\Rey} {\slsB}_1 +
	    {\slsB}_2 + {\rm i}(\kappa l c_z + n_s n c_{\theta}) {\slsA} -
	    {\Diff}_\bita \bitb(\bita_{\rm tw}) \right) \bitv_\bita + \bitC v_f  &=& \lambda \slsA \bitv_\bita \\
    \bitQ \bcdot \bitv_\bita &=& 0 
  \end{array}
  \right.
  \label{eq:eigprob}.
\end{equation}
System (\ref{eq:eigprob}) constitutes a constrained generalised
eigenproblem which can be expressed in matrix form as
\begin{equation}
  \sbA \sbx = \lambda \sbM \sbx
  \label{eq:ceigprob},
\end{equation}
where
\begin{equation}
  \begin{array}{c}
    \sbA = \left[
      \begin{array}{cc}
	\frac{1}{\Rey} {\slsB}_1 + {\slsB}_2 + {\rm i}(\kappa l c_z +
	n_s n c_{\theta}) {\slsA} - {\Diff}_\bita \bitb(\bita_{\rm
	  tw}) & \bitC \\
	\bitQ^{T} & 0
      \end{array}
      \right], \\
    \sbM = \left[
      \begin{array}{cc}
	\slsA & {\bf 0} \\
	      {\bf 0}^{T} & 0
      \end{array}
      \right]
    \qquad \textrm{and} \qquad
    \sbx = \left(
      \begin{array}{c}
	\bitv_a \\
	v_f
      \end{array}
      \right).
  \end{array}
  \label{eq:MAx}
\end{equation}

Since we are only interested in leading eigenvalues (those with larger
real part) while usual iterative methods are good at capturing
dominant eigenvalues (those with larger modulus), we solve
(\ref{eq:ceigprob}) in shift-invert mode:
\begin{equation}
  (\sbA - \sigma \sbM)^{-1} \sbM \sbx = \nu \sbx, \qquad
  \nu=\frac{1}{\lambda-\sigma}
  \label{eq:SIeigprob},
\end{equation}
with a real shift $\sigma$ applied to avoid the neutrally stable
double eigenvalue at the origin corresponding to infinitesimal
rotations and translations.

The method used is Arnoldi iteration with Krylov-based iteration for
the solution of the linear system involved at each step
\cite[][]{Freitag_PhD_07}. We use the robust routines by
\cite*{ARPACK_96}. Stokes preconditioning is employed as for the
Newton method to speed up convergence.

\subsection{Normal form reduction}



We will be concerned here with the stability and bifurcation of
travelling-wave-type solutions. Strictly speaking, a Poincar\'e
section should be defined and the bifurcation analysis carried out for
the map defined by the resulting Poincar\'e application. In this
setting, travelling waves would be seen as fixed points and modulated
waves as periodic orbits. It is common practice, though, to
approximate the map to any desired accuracy with a system of
differential equations \cite[see][chapter 4]{Guckenheimer_B_83}.
Bifurcations of maps can thus be studied by constructing
continuous-time normal forms. This comes out naturally when studying
the stability of relative equilibria in a comoving reference frame
(\ref{eq:infpert}). In the close neighbourhood of the relative (or
degenerate) equilibrium, such an analysis is perfectly accurate
\cite*[][]{Rand_ARMA_82,Krupa_SJMA_90}. These travelling waves may
typically undergo Hopf bifurcations. The bifurcation is considered a
Hopf bifurcation from a relative equilibrium rather than a
Neimark-Sacker (Hopf bifurcation for maps) because of the special type
of time-dependence involved. The bifurcating quasi-periodic solution
is a modulated travelling wave which is just a periodic orbit in the
comoving frame in which the travelling wave is stationary. Modulated
waves that bifurcate via Hopf bifurcations from drifting waves in
systems with $SO(2)$ symmetry were classified in terms of the spatial
symmetry of the drifting wave by \cite{Rand_ARMA_82}. The
classification was completed by \cite*{GoLeMe_JNS_00} to include the
spatiotemporal symmetry of the modulated wave. The appropriate way to
investigate these solutions is to study the dynamics normal to the
group orbit \cite[][]{Krupa_SJMA_90}, which in our case is equivalent
to considering the dynamics in a moving reference frame in which the
travelling wave is stationary.

A problem arises when, for given values of the parameters, coexistence
of several waves occurs and, possibly, of one or more modulated
waves. In this case, each wave will generally drift with its own phase
speed, so that there is no unique moving reference frame in which
their degeneracy could be lifted simultaneously. This is easily
overcome by stating the problem back with the use of Poincar\'e
sections, combined with continuous-time interpolation of the
corresponding maps. This of course only solves the problem as long as
the limits of validity of the interpolation are not infringed. In
particular, special care should be taken when studying the dynamics in
the vicinity of homoclinic orbits, where arbitrarily small
perturbations can lead to qualitatively different behaviour due to
manifolds entanglement. In the absence of translational symmetry,
complex dynamics would arise, \ie horseshoes
\cite[][]{Smale_BAMS_67,Guckenheimer_B_83}.
However, it is possible to show that, due to the translational
symmetry of the pipe, if a point lies on both the stable and the
unstable manifold of a saddle such as a lower-branch travelling wave,
then the unstable manifold is contained within the stable manifold and
complex dynamics does not occur \cite*[][]{ChossatLauterbach_B_00}.

It will be argued that, in the problem at hand, there are only two
active modes in the range of parameters studied, all other modes being
slaved to these two via the centre manifold theorem. This, added to
the translational symmetry of the pipe, results in the ability of
capturing the dynamics within this region of parameter space with a
pair of amplitude equations:
\begin{equation}
  \left\{
  \begin{array}{l}
    \dot{x}=F_1(x,y),\\
    \dot{y}=F_2(x,y),
  \end{array} \right.
  \label{eq:AmpEq}
\end{equation}
where $x$ and $y$ are real amplitudes related to velocity fields in
(\ref{eq:NS}). The simplest expansion for $F_1$ and $F_2$ that is
compatible with the dynamics observed upon integration of
(\ref{eq:ODEs}) is the subject of normal form theory and will be
developed along the analysis.

\section{Results}\label{sec:results}

\subsection{Travelling waves}


The special stability properties of the specific family of travelling
waves in the $2$-fold periodic subspace singles them out for the kind
of analysis performed in this study. These waves, whose existence
relies on the {\em Self-Sustaining Process} (SSP)
\cite*[][]{Waleffe_SAM_95,Wal_PoF_97}, were first computed by
\cite{FaEc_PRL_03} and \cite{WeKe_JFM_04} using volume forcing
homotopy. They are characterised by discrete $2$-fold azimuthal
periodicity ($Z_{2}$, \ie $n_s=2$) and a combined shift-reflect
symmetry ($S$).

\begin{figure}
  \begin{center}
    \begin{tabular}{cccc}
      $z=0$ & $z=\sLambda/6$ & $z=\sLambda/3$ & $z=\sLambda/2$ \\
      \includegraphics[width=0.23\linewidth,clip]{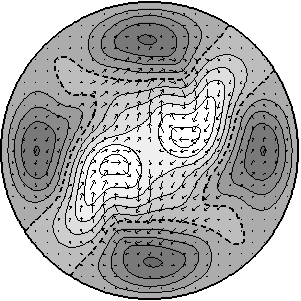} &
      \includegraphics[width=0.23\linewidth,clip]{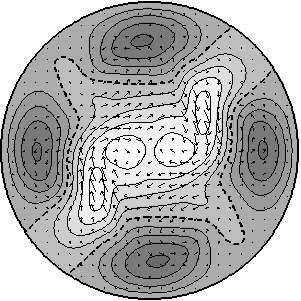} &
      \includegraphics[width=0.23\linewidth,clip]{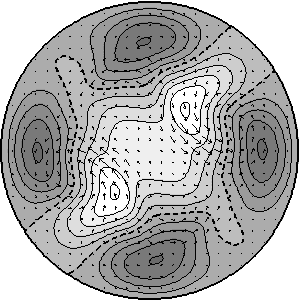} &
      \includegraphics[width=0.23\linewidth,clip]{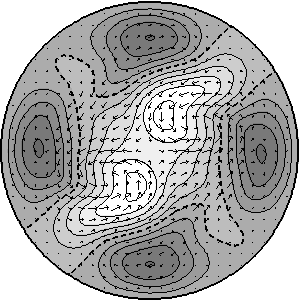}\\
      \includegraphics[width=0.23\linewidth,clip]{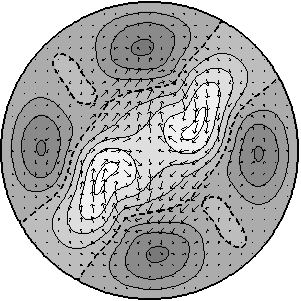} &
      \includegraphics[width=0.23\linewidth,clip]{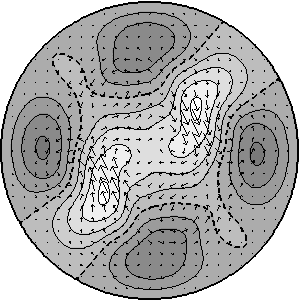} &
      \includegraphics[width=0.23\linewidth,clip]{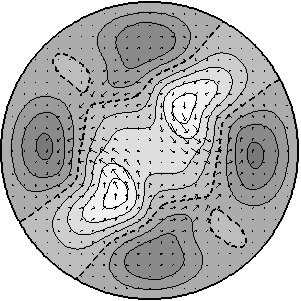} &
      \includegraphics[width=0.23\linewidth,clip]{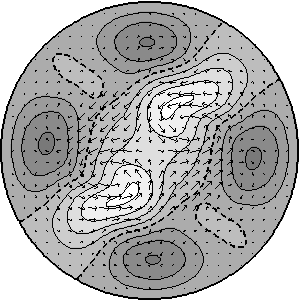}\\
    \end{tabular}
  \end{center}
  \caption{Cross-sections spanning half a wavelength
    ($\sLambda/2=\upi/\kappa$) of the upper-branch (top row) and the
    lower-branch (bottom row) travelling waves for $\kappa=1.52$,
    $\Rey=1600$. Axial perturbation velocity contour-lines are
    equispaced with intervals of $\Delta u_z=0.1 U$. Regions where the
    axial velocity is faster or slower than the laminar parabolic
    profile are coded in dark and light, respectively, with the zero
    contour-line marked with a dashed line. In-plane velocity vectors
    are also shown.}
  \label{fig:TWCS}
\end{figure}

For a given $\kappa$ and above a certain critical $\Rey$, they appear in
pairs, one that is closer to the laminar solution, the other being
further away in all respects and whatever the property examined
(friction factor, energy contents, distance in phase space, etc). They
are accordingly tagged as a lower-branch and an upper-branch solution,
respectively.  Figure~\ref{fig:TWCS} (see online movies) shows axial
velocity contours together with in-plane velocity vectors on four
cross-sections spanning half a wavelength for $\kappa=1.52$,
$\Rey=1600$ upper-branch (top row) and lower-branch (bottom row)
travelling waves. The discrete $2$-fold azimuthal periodicity can be
seen in all cross-sections.  Also the combined shift-reflect symmetry
can be grasped by comparing the $z=0$ and $z=\sLambda/2$
cross-sections of either travelling wave. The reflection plane is
tilted with $\theta_0 \simeq \upi/4$. The complementary
half-wavelength is not shown, as it can be constructed by mere
reflection of the given cross-sections.
The travelling waves shown are representative of the full family in
that they are characterised by wobbling low-speed streaks (light
shading) in the pipe central region and fairly streamwise-independent
high-speed streaks in the wall region (dark). Leaving aside weak
secondary vortices, a couple of strong vortex pairs can be identified
from the velocity vectors. In each of the pairs, the vortices are not
of equal intensity, but alternate in strength along the axial
direction. Except for the higher axial velocities and the slightly
wider low speed streaks, not much distinguishes the upper-branch
solution from the lower-branch solution.

\begin{figure}
  \begin{center}
    \begin{tabular}{cccccc}
      \raisebox{0.24\linewidth}{(\textit{a})} \raisebox{0.015\linewidth}{\parbox{.4cm}{\vspace{-2.7cm}$0$\\  \vspace{2.1cm}\\ $\sLambda$}} \hspace{-0.3cm} &
      \includegraphics[width=0.35\linewidth,clip]{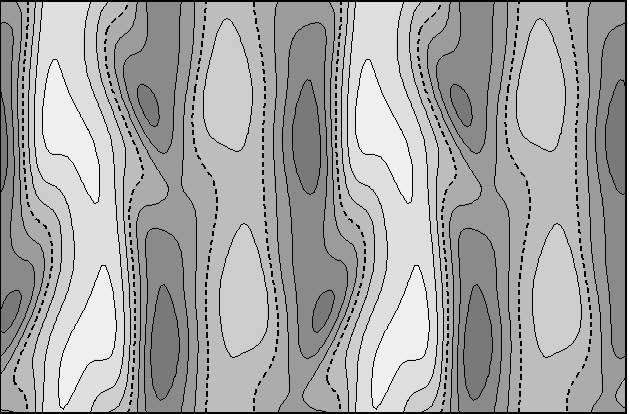} &
      \raisebox{0.24\linewidth}{(\textit{b})} \raisebox{0.015\linewidth}{\parbox{.4cm}{\vspace{-2.7cm}$0$\\  \vspace{2.1cm}\\ $\sLambda$}} \hspace{-0.3cm} &
      \includegraphics[width=0.35\linewidth,clip]{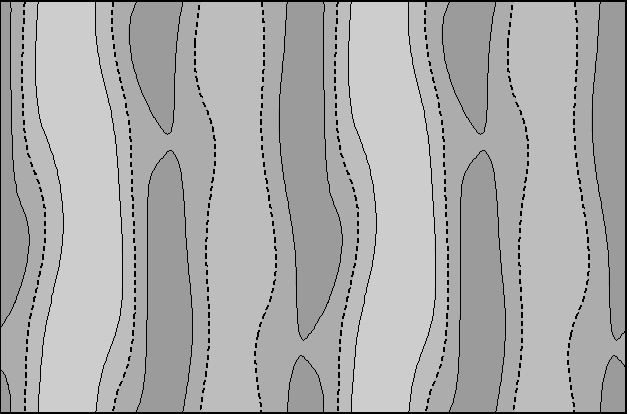} \\
      ~ &
      $0$ \hfill $2 \upi$ &
      ~ &
      $0$ \hfill $2 \upi$ \\
    \end{tabular}
  \end{center}
  \caption{Radial sections at $r=0.65$ of (\textit{a}) the
    upper-branch and (\textit{b}) the lower-branch travelling waves
    for $\kappa=1.52$, $\Rey=1600$. Radial velocity contour-lines are
    equispaced with intervals of $\Delta u_r=0.0072 U$. Regions where
    the flow goes towards the walls or the pipe center are shaded dark
    and light, respectively.}
  \label{fig:TWRS}
\end{figure}

Radial-velocity contour-lines on a radial plane at $r=0.65$ have been
plotted in figure~\ref{fig:TWRS} for both travelling waves.
The difference between upper-branch and lower-branch waves becomes
clearer, the former involving much larger radial transport. These
sections display clearly both the azimuthal periodicity and the
shift-reflect symmetry of the waves.
Structures resembling these and other travelling waves have been
observed experimentally in developed turbulence
\cite[][]{HVWNFEWKW_SCI_04} as well as in numerical simulation
\cite*[][]{KeTu_JFM_07,ScEcVo_PRE_07,WiKe_PRL_08}.


They appear in saddle-node bifurcations starting from $\Rey \simeq
1358.5$ for the optimal wave number $\kappa \simeq 1.55$
\cite[][]{WeKe_JFM_04}. Lower-branch solutions exhibit a single
unstable eigenmode when considered in the azimuthal subspace where
they live, which makes them accessible through time integration if
appropriately combined with edge tracking techniques.
In this sense, they are edge states \cite[][]{ScEcYo_PRL_07} of the
$2$-fold azimuthally-periodic pipe, and they constitute an example of
non-uniqueness of the edge state, since they coexist as attractors
within the critical threshold, and for some range of $\kappa$ and
$\Rey$, with another family of travelling waves that are both
shift-reflect and mirror symmetric and that have also a single
unstable direction \cite[][]{DuWiKe_JFM_08}.
The stable manifold of these waves separates, albeit only locally,
initial conditions that decay uneventfully from others that undergo
turbulent transients. For this reason, lower-branch solutions are
believed to play a central role in the transition process.

\begin{figure}
  \begin{center}
    \begin{tabular}{cccc}
      \raisebox{0.33\linewidth}{(\textit{a})}\hspace{-0.3cm} & \includegraphics[height=0.32\linewidth,clip]{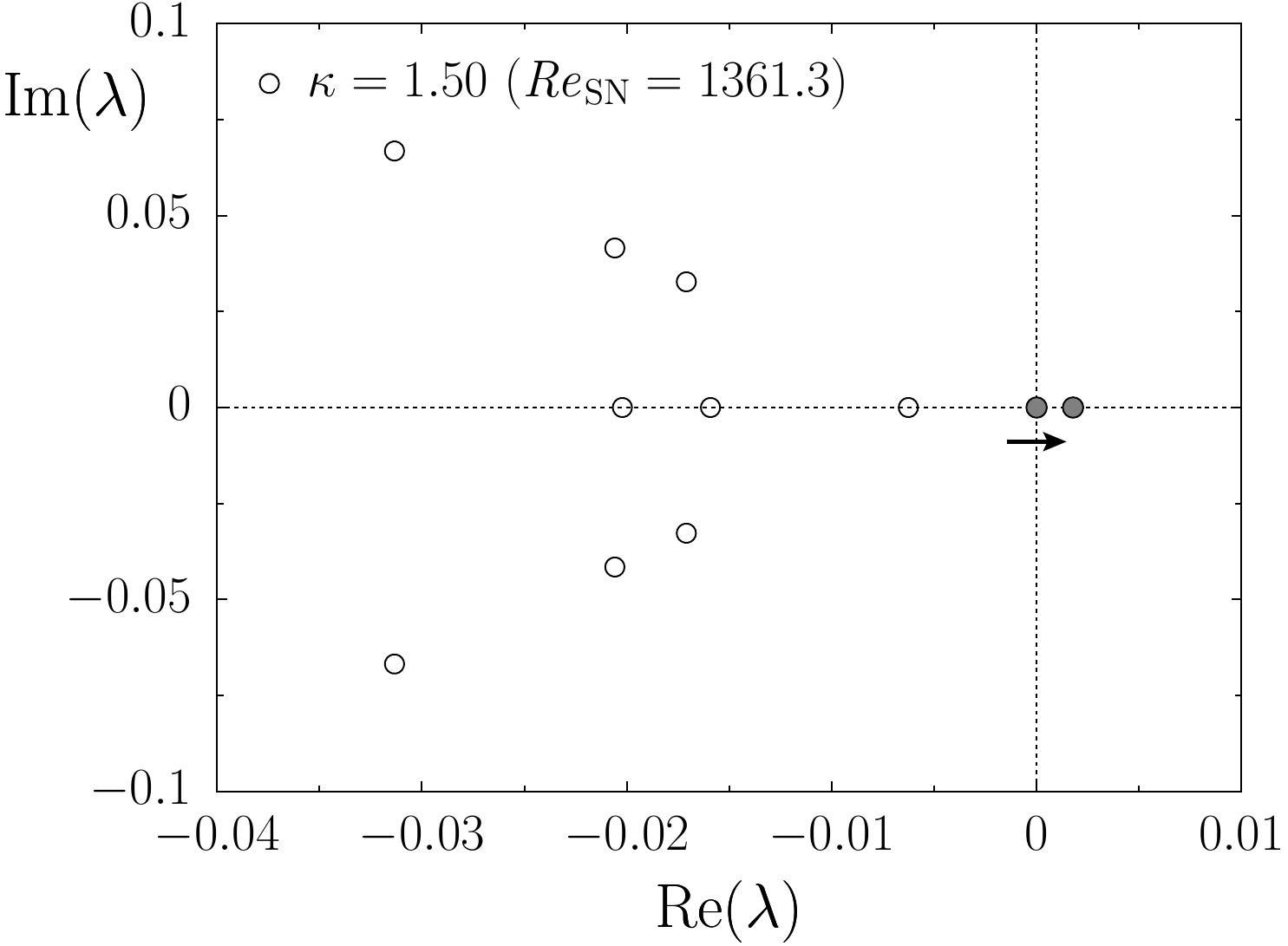} &
      \raisebox{0.33\linewidth}{(\textit{c})}\hspace{-0.3cm} & \includegraphics[height=0.32\linewidth,clip]{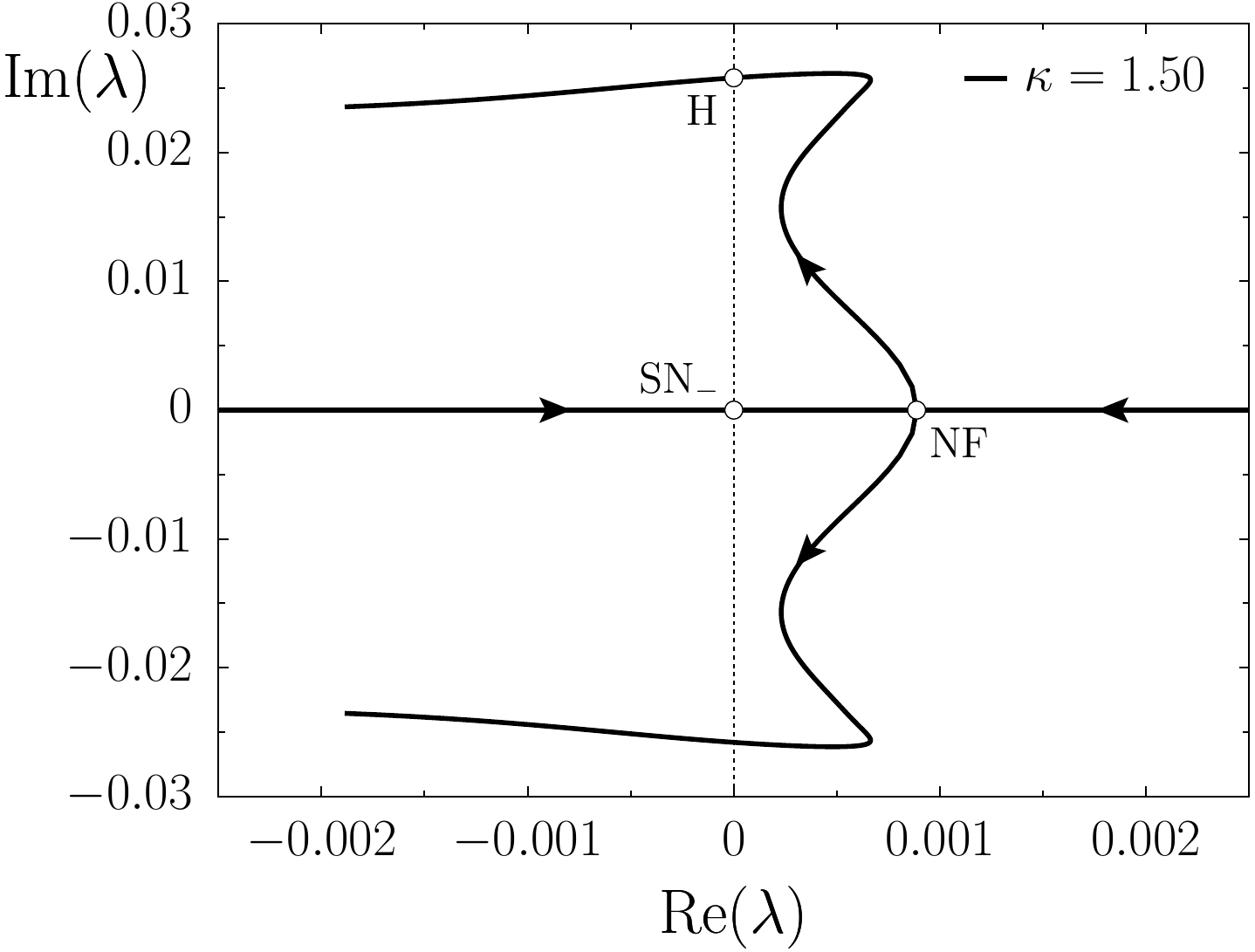}\\
      \raisebox{0.33\linewidth}{(\textit{b})}\hspace{-0.3cm} & \includegraphics[height=0.32\linewidth,clip]{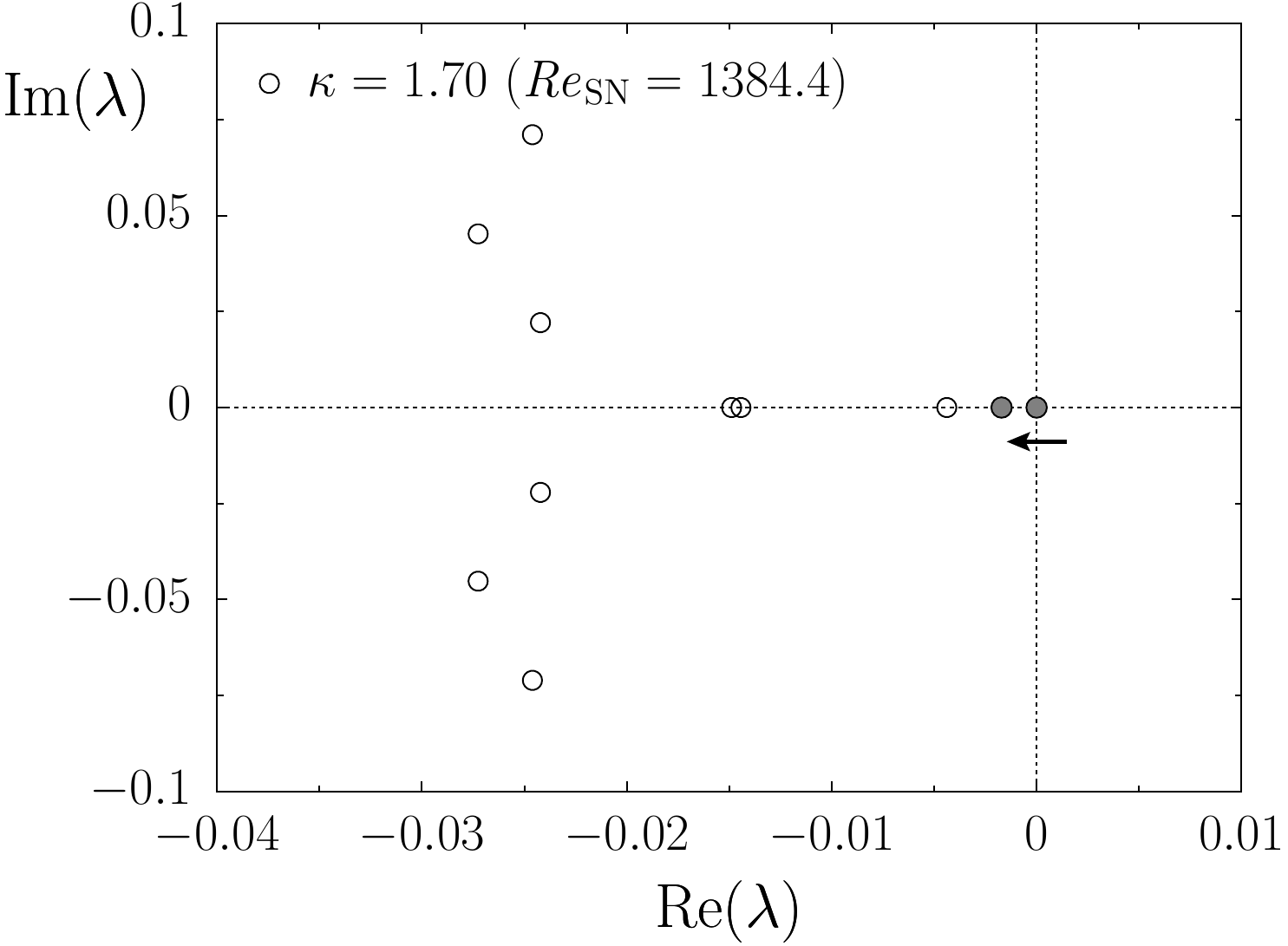} &
      \raisebox{0.33\linewidth}{(\textit{d})}\hspace{-0.3cm} & \includegraphics[height=0.32\linewidth,clip]{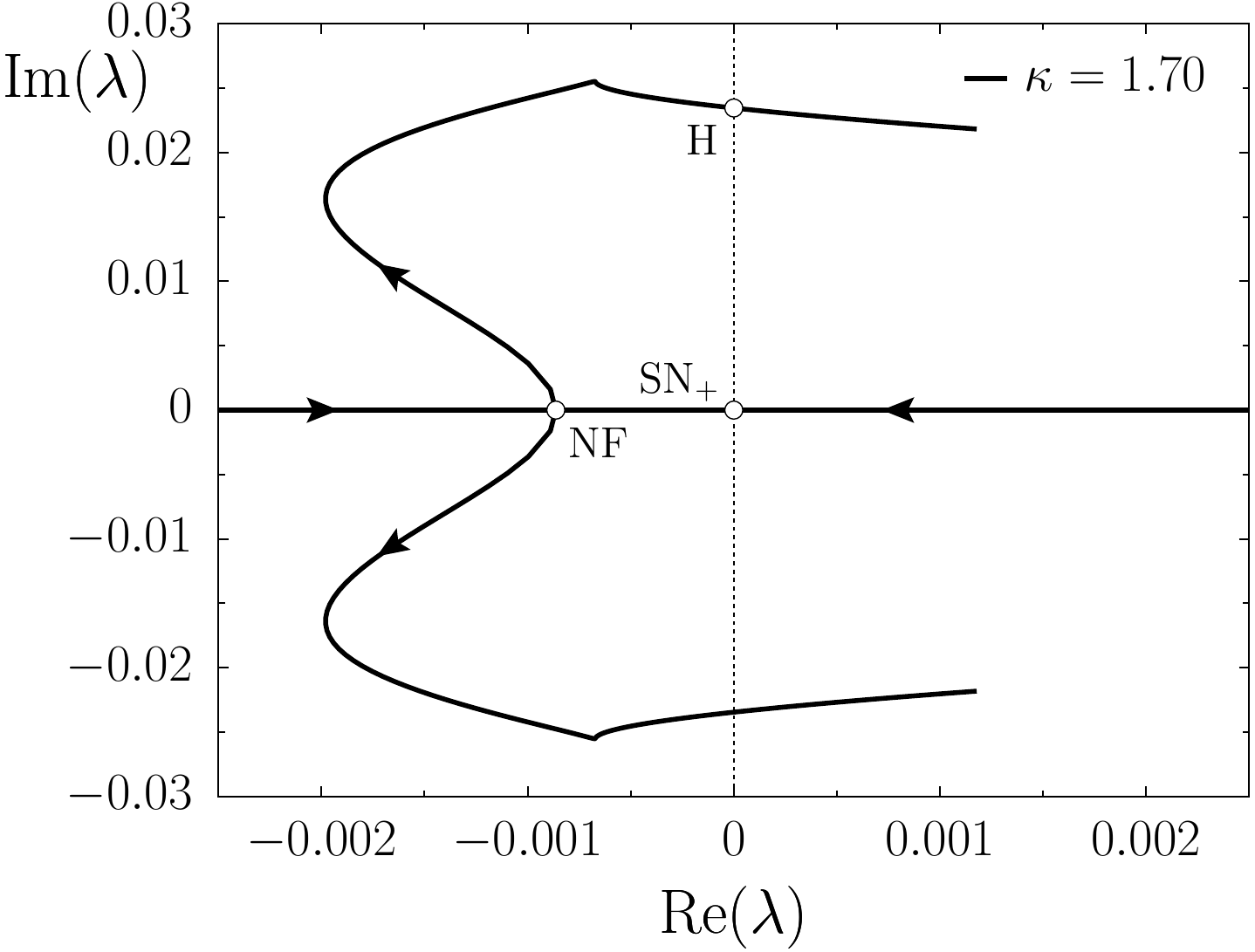}
    \end{tabular}
  \end{center}
  \caption{Rightmost eigenvalues of (\textit{a}) $\kappa=1.50$ and
    (\textit{b}) $\kappa=1.70$ travelling waves at their respective
    saddle-node bifurcation ($\rm{SN}$). The direction of the
    zero-crossing from lower to upper branch is indicated with an
    arrow. The eigenvalues marked as filled circles have been followed
    across the saddle-node point and beyond for both (\textit{c})
    $\kappa=1.50$ and (\textit{d}) $\kappa=1.70$. Saddle-node
    ($\rm{SN}$), node-focus ($\rm{NF}$) and Hopf ($\rm{H}$) points are
    identified with circles.}
  \label{fig:Spectra}
\end{figure}

As already pointed out, lower-branch solutions, when continued down in
$\Rey$, merge with upper-branch solutions in a saddle-node-type
bifurcation. Around the saddle-node point, a real eigenvalue must
cross the imaginary axis, either adding or subtracting an unstable
direction. As a matter of fact, both scenarios occur and the stability
properties of upper-branch solutions depend on the axial wavenumber
$\kappa$. Upper-branch long waves are unstable, whilst short waves are
stable. The spectrum of a long ($\kappa=1.50$) and of a short
($\kappa=1.70$) wave are shown in
figures~\ref{fig:Spectra}(\textit{a}) and
\ref{fig:Spectra}(\textit{b}), respectively, at their respective
saddle-node points. The arrow below the bifurcating eigenvalue points
in the crossing direction when followed from lower to upper
branch. For $\kappa=1.50$, the solution goes from one to two
unstable directions, while for $\kappa=1.70$ it goes from one to nil.
Figures~\ref{fig:Spectra}(\textit{c}) and
\ref{fig:Spectra}(\textit{d}) show the trajectories followed by the
two rightmost eigenvalues (filled circles in
figures~\ref{fig:Spectra}\textit{a},\textit{b}) when continued across
the saddle-node ($\rm{SN}$) and beyond, for $\kappa=1.50$ and
$\kappa=1.70$, respectively. In both cases, the eigenvalues collide
and go complex in a node-focus ($\rm{NF}$) transition, but they can do
so at either side of the imaginary axis. When followed further to
higher $\Rey$ along the upper branch, long and short solutions undergo
a stabilising or destabilising Hopf bifurcation ($\rm{H}$),
respectively, implying the existence of modulated waves (relative
periodic orbits).

\begin{figure}
  \begin{center}
    \begin{tabular}{cccccccc}
      \raisebox{0.235\linewidth}{(\textit{a})}\hspace{-.6cm} &
      \includegraphics[height=0.23\linewidth,clip]{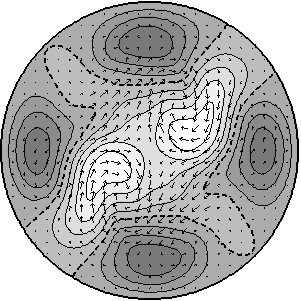} &
      \raisebox{0.235\linewidth}{(\textit{b})}\hspace{-.6cm} &
      \includegraphics[height=0.23\linewidth,clip]{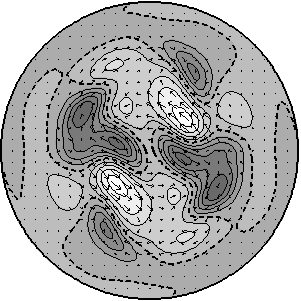} &
      \raisebox{0.235\linewidth}{(\textit{c})}\hspace{-.6cm} &
      \includegraphics[height=0.23\linewidth,clip]{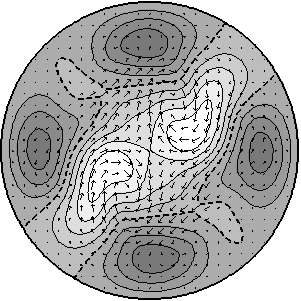} &
      \raisebox{0.235\linewidth}{(\textit{d})}\hspace{-.6cm} &
      \includegraphics[height=0.23\linewidth,clip]{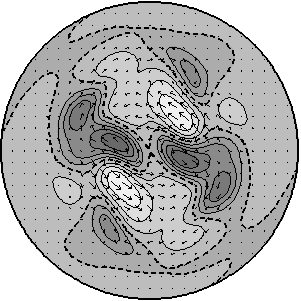}
    \end{tabular}
  \end{center}
  \caption{(\textit{a}) $\kappa=1.50$ and (\textit{c}) $\kappa=1.70$
    travelling waves at their respective saddle-node bifurcation
    points ($\rm{SN}$ in figure~\ref{fig:SNCS}). Also shown are the
    bifurcating eigenfunctions for (\textit{b}) $\kappa=1.50$ and
    (\textit{d}) $\kappa=1.70$. Axial perturbation velocity
    contour-lines are equispaced in intervals of $\Delta u_z=0.1 U$.}
  \label{fig:SNCS}
\end{figure}

Cross-sections of both travelling waves at their respective
saddle-node points have been plotted in
figures~\ref{fig:SNCS}(\textit{a}) and \ref{fig:SNCS}(\textit{c}),
together with their corresponding crossing eigenfunctions
(figures~\ref{fig:SNCS}\textit{b} and \ref{fig:SNCS}\textit{d}).
It is clear from the contour plots that the effect of varying $\kappa$
on the solutions does not affect the flow structure dramatically,
aside from the stretching and the stability properties.
It is also remarkable how the eigenfunctions corresponding to two
different eigenvalues
(figures~\ref{fig:SNCS}\textit{c},\textit{d}) look so similar to
each other, even for different $\kappa$. This is a signature of the
tendency to become parallel, which ultimately leads to collision in a
node-focus ($\rm{NF}$) transition. Both eigenfunctions preserve the
same symmetries as the travelling waves, so that all solutions arising
from their bifurcation must necessarily possess the same symmetries.

Wavenumber continuity suggests that there might be an intermediate
value of $\kappa$ for which the node-focus point can be brought to the
origin of the complex plane. Appropriate tuning of $\kappa$ and $\Rey$
would therefore help unfold a codimension-two double zero bifurcation,
a complete study of which is due to \cite*{Takens_IHES_74} and to
\cite*{Bogdanov_FAA_75}.  As a matter of fact, the trajectories of the
complex eigenvalues, following a left opening parabola plus an
eventual right turn, indicate that for a certain range of $\kappa$ an
additional pair of crossings will take place, thus implying additional
Hopf bifurcations closely related to the appearance of periodic orbits
in a double-zero bifurcation.

\subsection{Modulated travelling waves}

\begin{figure}
  \begin{center}
    \begin{tabular}{ccccc}
      \raisebox{0.32\linewidth}{(\textit{a})}\hspace{-0.5cm} &
      \includegraphics[height=0.3\linewidth,clip]{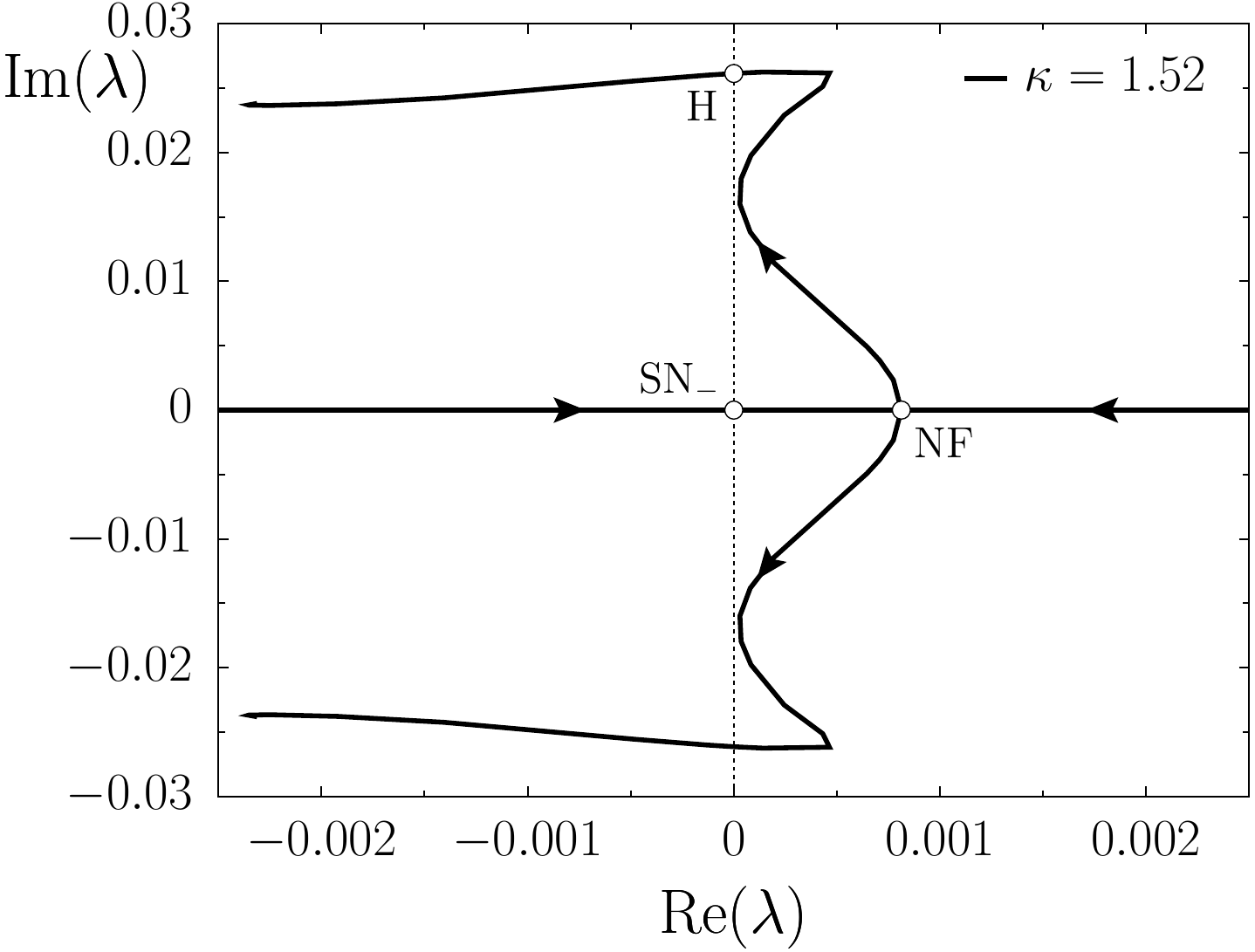} &
      \raisebox{0.32\linewidth}{(\textit{b})}\hspace{-0.5cm} &
      \raisebox{0.04\linewidth}{\includegraphics[height=0.25\linewidth,clip]{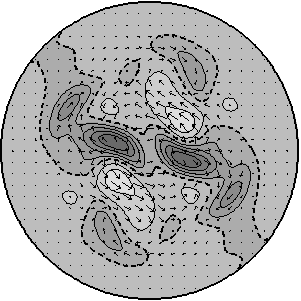}} &
      \raisebox{0.04\linewidth}{\includegraphics[height=0.25\linewidth,clip]{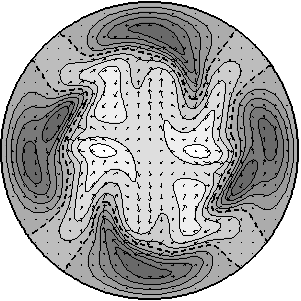}}
    \end{tabular}
  \end{center}
  \caption{(\textit{a}) Trajectories of the two rightmost eigenvalues
    for $\kappa=1.52$.  (\textit{b}) Real (left) and imaginary (right)
    parts of the corresponding eigenvectors at the Hopf bifurcation
    ($\rm{H}$, $\Rey=1775.4$).}
  \label{fig:EV1.52}
\end{figure}

Figure~\ref{fig:EV1.52}(\textit{a}) depicts the trajectories of the
relevant eigenvalues for $\kappa=1.52$. They are extremely close to
those observed for $\kappa=1.50$, except that the nose is getting
tangent to the imaginary axis. The implications of this will be
discussed later on. Let us focus on the Hopf bifurcation ($\rm{H}$)
for the time being. The crossing takes place at $\Rey=1775.4$ on the
upper branch, with the arrows pointing towards higher $\Rey$.
The real and imaginary parts of the bifurcating eigenfunction are
shown in figure~\ref{fig:EV1.52}(\textit{b}) only for completeness,
since it is difficult to extract from them relevant information on the
nature of the instability that they carry. As already noted, the
instability does not break any of the symmetries exhibited by the
bifurcating travelling wave.

The existence of a Hopf bifurcation implies the appearance of a
time-periodic solution, in this case a modulated wave or relative
periodic orbit. Since the crossing is leftwards as $\Rey$ is
increased, it necessarily carries with it a stabilisation of the
upper-branch travelling wave, which was unstable at onset. The Hopf
bifurcation happens to be supercritical in the sense that the first
Lyapunov coefficient is negative \cite[][]{Kuznetsov_B_95}. This we
infer from the fact that time evolution unveils a stable branch of
modulated waves pointing towards decreasing $\Rey$.

Simple time evolution starting from the $(\kappa,\Rey)=(1.52,1600)$
unstable upper-branch travelling wave (figure~\ref{fig:TWCS}, top row)
departs in an oscillatory fashion, as predicted by the complex pair of
unstable eigenvalues, to nonlinearly converge onto a modulated wave.

\begin{figure}
  \begin{center}
    \begin{tabular}{cccc}
      \raisebox{0.305\linewidth}{(\textit{a})}\hspace{-0.6cm} &
      \includegraphics[height=0.285\linewidth,clip]{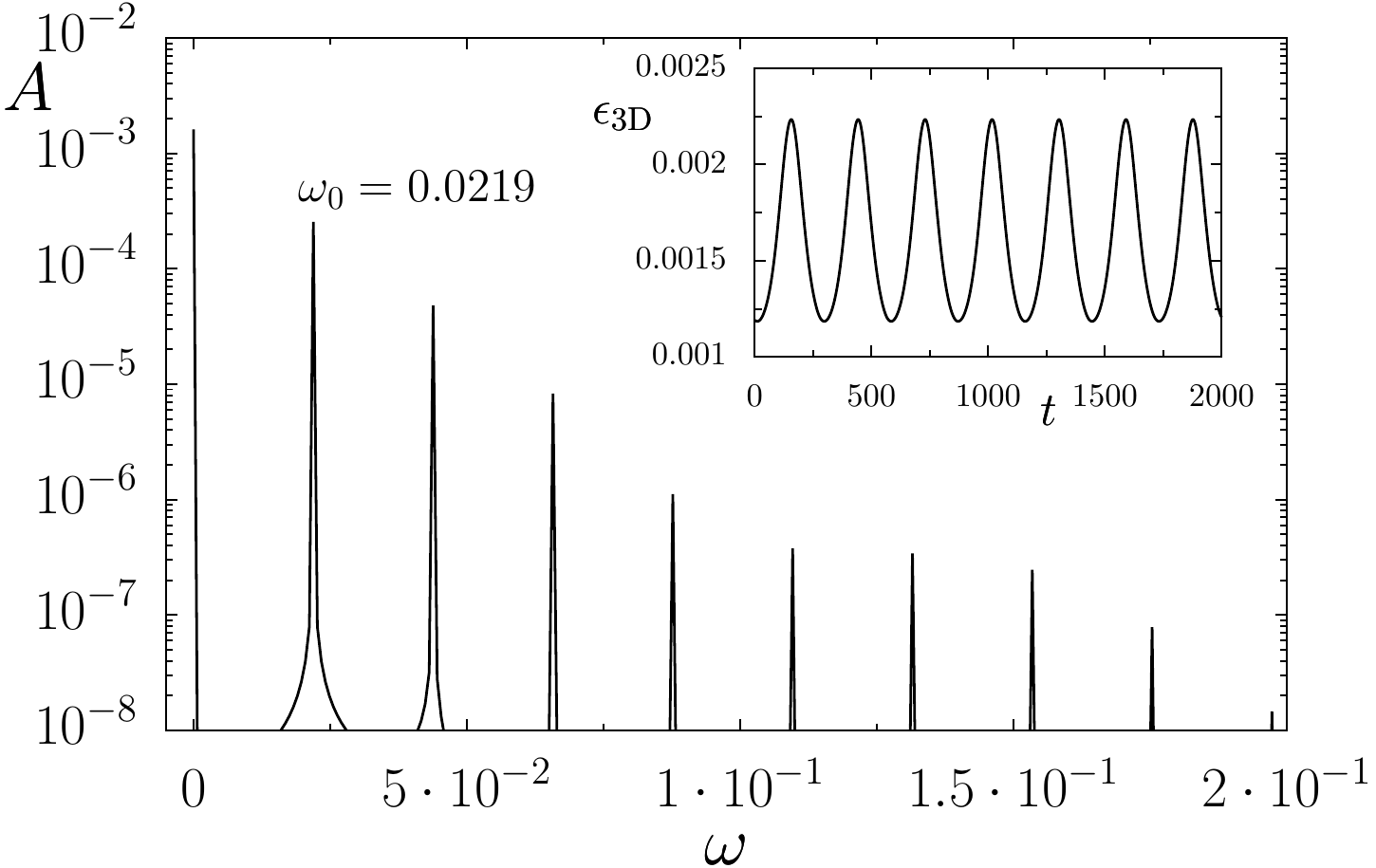} &
      \raisebox{0.305\linewidth}{(\textit{b})}\hspace{-0.6cm} &
      \includegraphics[height=0.27\linewidth,clip]{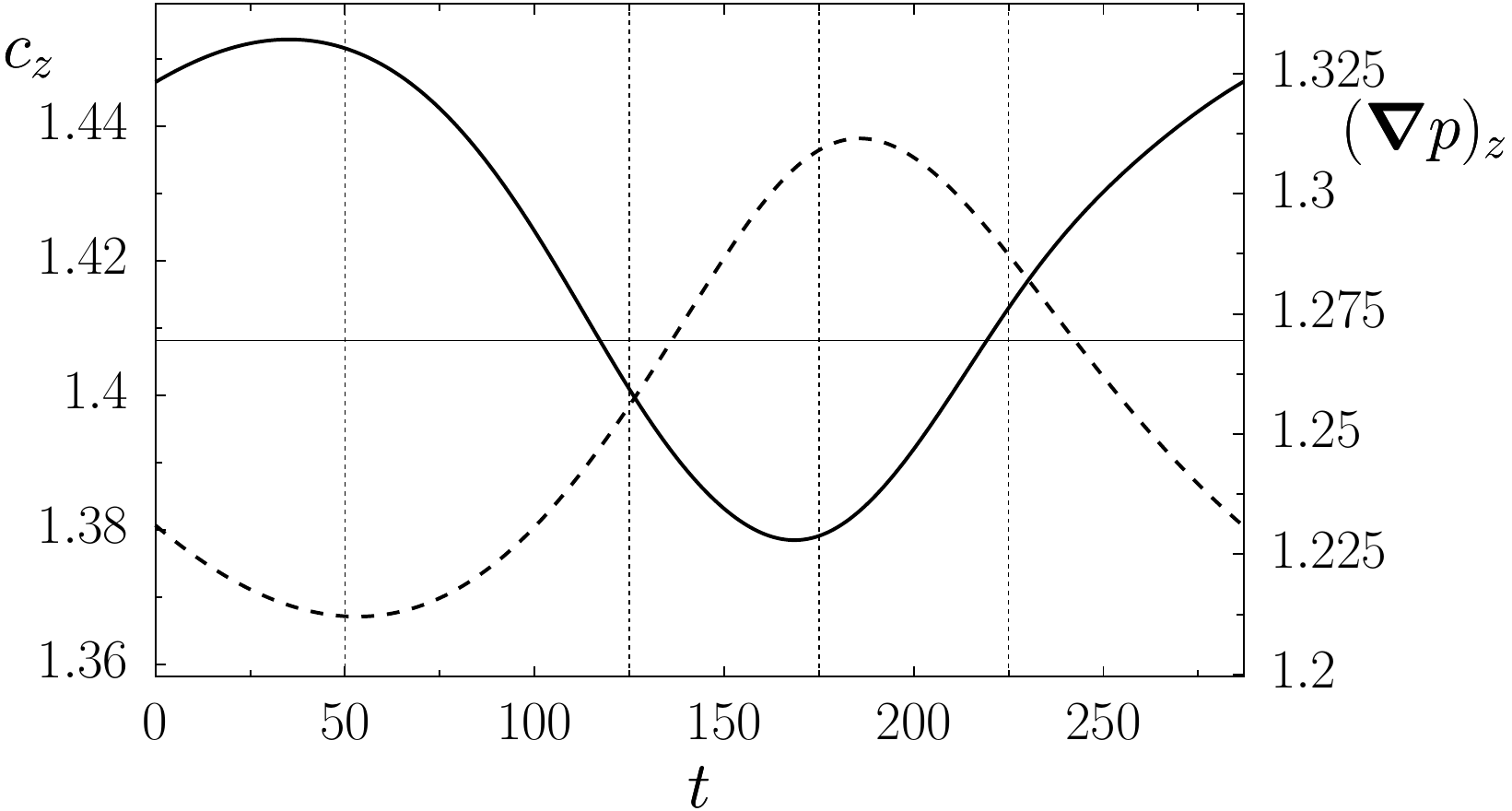}
    \end{tabular}
  \end{center}
  \caption{(\textit{a}) Fourier transform of the non-axisymmetric
    streamwise-dependent modal energy contents ($\varepsilon_{\rm
    3D}$) time-stamp for the $(\kappa,\Rey)=(1.52,1600)$ modulated
    wave. The time signal is plotted in the inset. (\textit{b})
    Advection speed ($c_z$, solid line) and mean axial pressure
    gradient normalised by its laminar value ($(\bnabla p)_z$, dashed
    line) along a full period. The values for the
    $(\kappa,\Rey)=(1.52,1600)$ upper-branch travelling wave are
    indicated by the horizontal line.}
  \label{fig:RPOk1.52TS}
\end{figure}

Travelling waves, which are degenerate or relative equilibria, and
solutions bifurcating from them have a 
pure frequency associated to
the advection speed (solid-body translation). Local quantities such as
point velocities or pressures necessarily reflect on this frequency.
However, since it reflects a neutral direction, it can be suppressed by
restating the problem
in a comoving frame. Global quantities such as modal energies or
volume averaged fields naturally overlook solid-body rotation and
translation, making them suitable for a decoupled analysis in the
direction orthogonal to the degenerate drift. Thus, as justified
before, the bifurcation analysis of travelling waves can be carried
out analogously to that of fixed points, as long as special care is
taken in the neighbourhood of homoclinic connections
\cite[][]{Rand_ARMA_82,GoLeMe_JNS_00}. The Hopf bifurcation adds a
modulational frequency to the pure translational frequency. As a
result, global quantities cease to be constant and oscillate with this
frequency. The Fourier transform of the energy contained in
non-axisymmetric streamwise-dependent modes ($\varepsilon_{\rm 3D}$)
has been plotted in figure~\ref{fig:RPOk1.52TS}(\textit{a}), with
the time signal shown in the inset frame. The spectrum reveals that
the solution has a strong mean component and a peak angular frequency
at $\omega_0=0.0219$, corresponding to a period
$T=2\upi/\omega_0=286.7$. This period is extremely long when compared
with the streamwise advection time-scale, which is of order $2 \upi
/(\kappa c_z) \sim 3$. The signal is not strictly sinusoidal and some
energy is spread among a number of harmonics of $w_0$.
A full period of the axial advection speed ($c_z$) and of the mean
axial pressure gradient normalised by its laminar value ($(\bnabla
p)_z$) has been plotted in
figure~\ref{fig:RPOk1.52TS}(\textit{b}). Both are global quantities
and, as expected, oscillate with the modulational frequency
$\omega_0$. The horizontal line indicates the constant value of these
quantities for the unstable upper-branch travelling wave from which
the modulated wave bifurcates. The modulated wave orbits in phase
space around the travelling wave, so that the instantaneous values of
the former oscillate around those of the latter. The differences
between time-averaged quantities and travelling wave values are purely
due to nonlinear effects, since for this $\Rey$ and $\kappa$ we are
already some distance away from the linear influence of the
instability.

\begin{figure}
  \begin{center}
    \begin{tabular}{ccccc}
       & $t=50$ & $t=125$ & $t=175$ & $t=225$ \\
      \raisebox{0.23\linewidth}{$\kappa=0$}\hspace{-1.1cm} &
      \includegraphics[width=0.23\linewidth,clip]{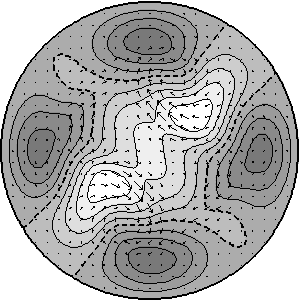} &
      \includegraphics[width=0.23\linewidth,clip]{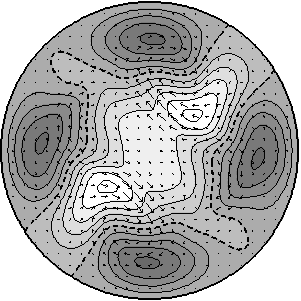} &
      \includegraphics[width=0.23\linewidth,clip]{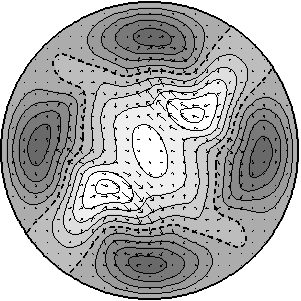} &
      \includegraphics[width=0.23\linewidth,clip]{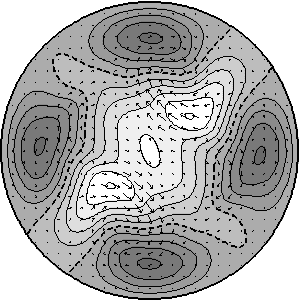}\\
      \raisebox{0.23\linewidth}{$\kappa=\sLambda/4$}\hspace{-1.1cm} &
      \includegraphics[width=0.23\linewidth,clip]{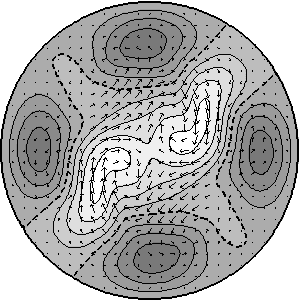} &
      \includegraphics[width=0.23\linewidth,clip]{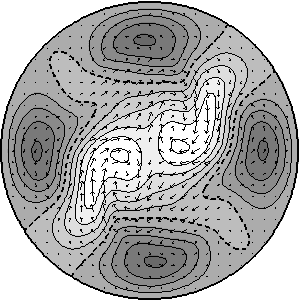} &
      \includegraphics[width=0.23\linewidth,clip]{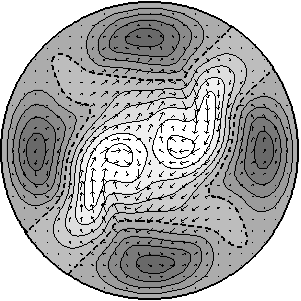} &
      \includegraphics[width=0.23\linewidth,clip]{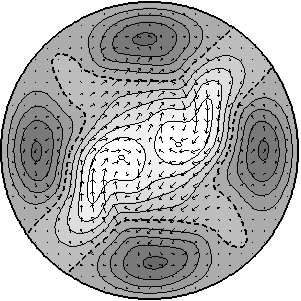}\\
    \end{tabular}
  \end{center}
  \caption{From left to right, $t=50$, $125$, $175$ and $225$
    snapshots of the ($(\kappa,\Rey)=(1.52,1600)$ modulated wave, as
    indicated by the vertical dotted lines in
    figure~\ref{fig:RPOk1.52TS}(\textit{b}). Shown are the $z=0$
    (top row) and $z=\sLambda/4$ (bottom row) cross-sections with
    axial velocity contour-lines equispaced in intervals of $\Delta
    u_z=0.1 U$. To avoid drift due to streamwise advection, snapshots
    are taken in a comoving frame travelling at the instantaneous
    advection speed from figure~\ref{fig:RPOk1.52TS}(\textit{b}).}
  \label{fig:RPOCS}
\end{figure}

Figure~\ref{fig:RPOCS} (see online movie) depicts a couple of
cross-sections ($z=0$, top row; $z=\sLambda/4$, bottom row) of the
modulated wave at four different time instants along one period (from
left to right, $t=50$, $125$, $175$ and $225$).
The modulation is obvious from the series of snapshots yet 
not
very prominent, as could be expected from the mild oscillation (within
$\pm 5\%$) of the time series in
figure~\ref{fig:RPOk1.52TS}(\textit{b}).
Comparison with figure~\ref{fig:TWCS} (top row) reveals that the
overall flow structure of the travelling wave is preserved on average
but is now dependent on time.

\begin{figure}
  \begin{center}
    \begin{tabular}{cccc}
      ~ & $t=50$ & ~ & $t=125$ \\
      \raisebox{0.015\linewidth}{\parbox{.4cm}{\vspace{-2.7cm}$0$\\  \vspace{2.1cm}\\ $\sLambda$}} \hspace{-0.3cm} &
      \includegraphics[width=0.35\linewidth,clip]{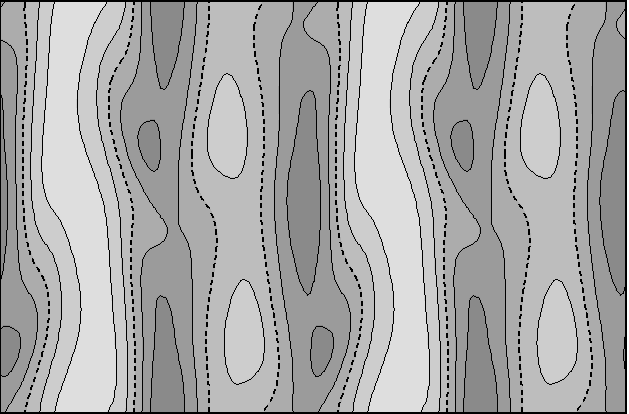} &
      \raisebox{0.015\linewidth}{\parbox{.4cm}{\vspace{-2.7cm}$0$\\  \vspace{2.1cm}\\ $\sLambda$}} \hspace{-0.3cm} &
      \includegraphics[width=0.35\linewidth,clip]{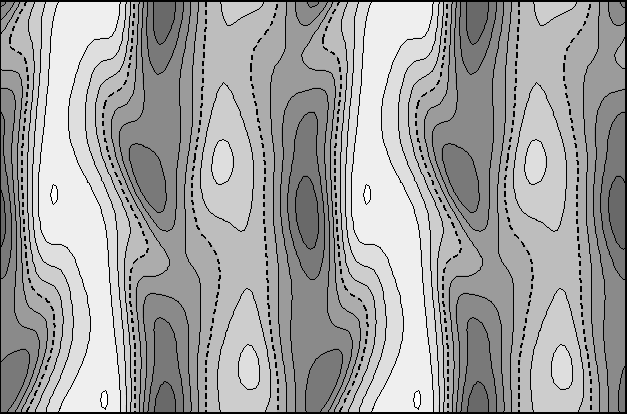} \\
      ~ &
      $0$ \hfill $2 \upi$ &
      ~ &
      $0$ \hfill $2 \upi$ \\
      ~ & $t=175$ & ~ & $t=225$ \\
      \raisebox{0.015\linewidth}{\parbox{.4cm}{\vspace{-2.7cm}$0$\\  \vspace{2.1cm}\\ $\sLambda$}} \hspace{-0.3cm} &
      \includegraphics[width=0.35\linewidth,clip]{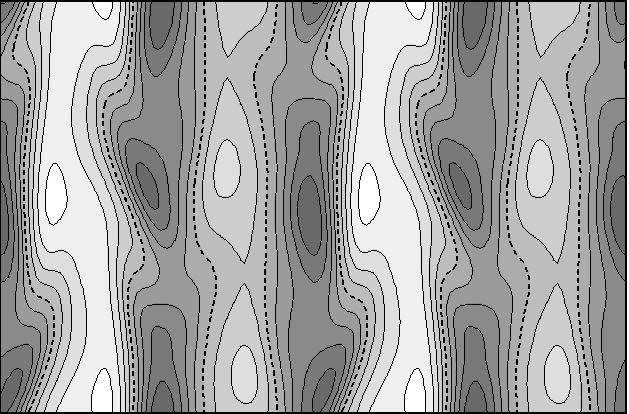} &
      \raisebox{0.015\linewidth}{\parbox{.4cm}{\vspace{-2.7cm}$0$\\  \vspace{2.1cm}\\ $\sLambda$}} \hspace{-0.3cm} &
      \includegraphics[width=0.35\linewidth,clip]{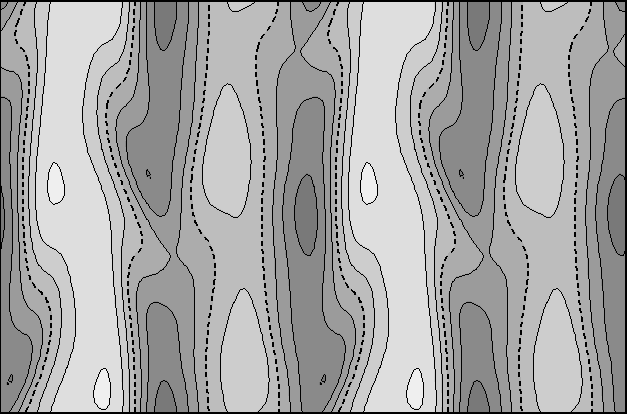} \\
      ~ &
      $0$ \hfill $2 \upi$ &
      ~ &
      $0$ \hfill $2 \upi$ \\
    \end{tabular}
  \end{center}
  \caption{Radial sections at $r=0.65$ of the
    $(\kappa,\Rey)=(1.52,1600)$ modulated wave. Snapshots are taken in
    a reference frame moving at the wave speed. Radial velocity
    contour-lines are equispaced with intervals of $\Delta u_r=0.0072
    U$. Regions where the flow goes towards the walls or the pipe
    center are shaded dark and light, respectively.}
  \label{fig:RPORS}
\end{figure}

Radial sections showing radial velocity contours have been plotted in
figure~\ref{fig:RPORS} (see online movie) to convey in a clear way the
modulation of the wave along a full period.
It also confirms that all symmetries are preserved. It will soon
become clear that the snapshot at $t=50$ is about the closest approach
to the lower-branch travelling wave, while all of them are
approximately equally distant from the upper-branch wave.

\begin{figure}
  \begin{center}
    \includegraphics[height=0.45\linewidth,clip]{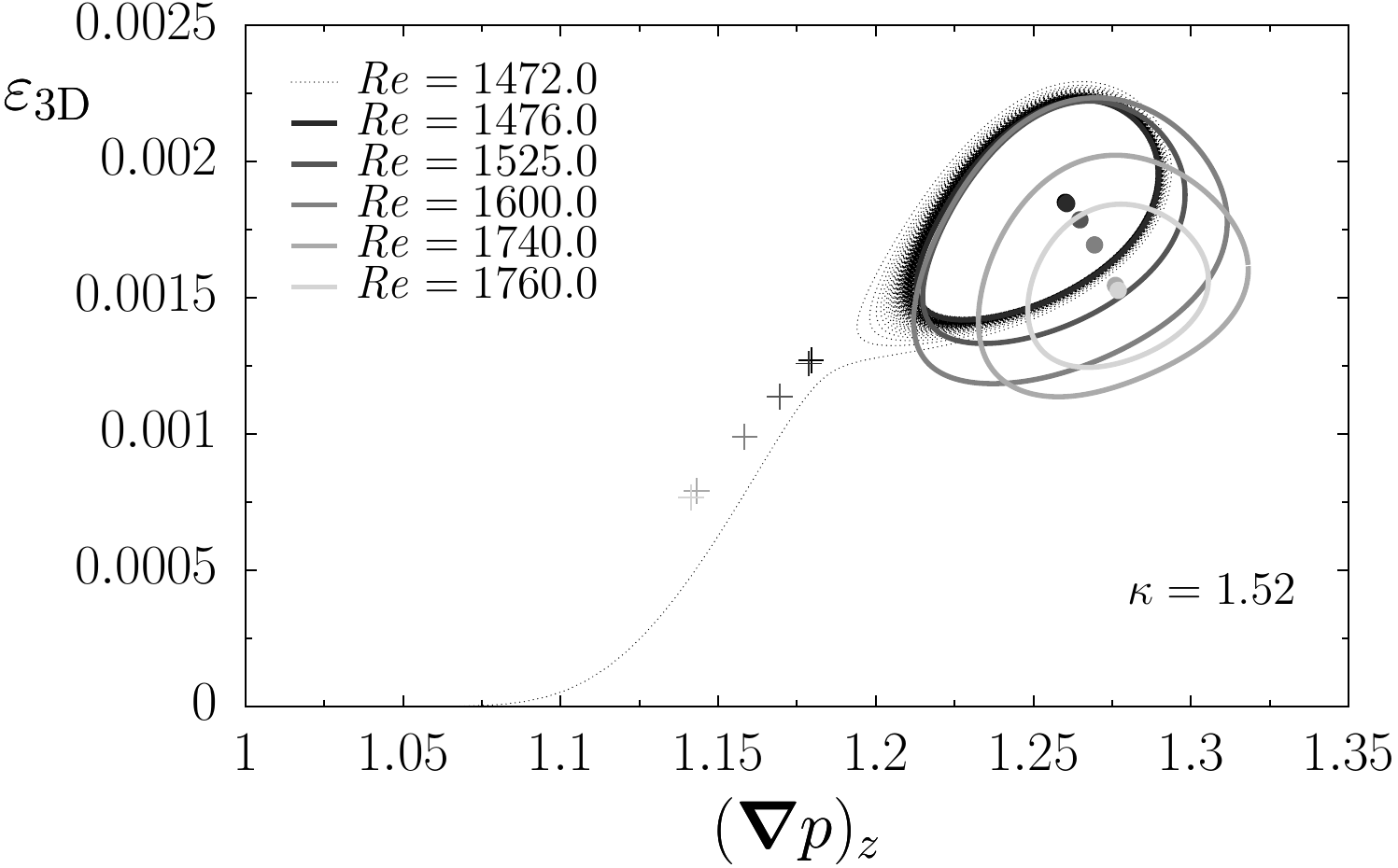}
  \end{center}
  \caption{Phase map projections of travelling and modulated waves
    along a $\Rey$-continuation for $\kappa=1.52$. The axes represent
    non-axisymmetric streamwise-dependent energy ($\varepsilon_{\rm
    3D}$) against mean axial pressure gradient ($(\bnabla
    p)_z$). Different gray intensities correspond to different $\Rey$
    as explained in the legend. Lower- and upper-branch travelling
    waves are marked with pluses and filled circles, respectively;
    modulated waves with solid lines and escaping trajectories with
    dotted lines.}
  \label{fig:RPOPhMap}
\end{figure}

Straightforward $\Rey$-continuation of the $\kappa=1.52$ modulated
waves by time evolution reveals that they bifurcate off the travelling
waves upper branch precisely at the Hopf bifurcation point at
$\Rey=1775.4$ ($\rm{H}$ in figure~\ref{fig:EV1.52}\textit{a}).
Figure~\ref{fig:RPOPhMap} shows phase map representations of the
modulated waves for a discrete set of $\Rey$ along the branch. Two
global quantities, $\varepsilon_{\rm 3D}$ and $(\bnabla p)_z$, have
been chosen to get rid of axial drift effects.
Increasing $\Rey$ from $\Rey=1600$ (towards lighter gray), the
modulated wave (solid closed loops) shrinks and closes on the
upper-branch travelling wave (filled circles), to end up merging at
the Hopf point ($\Rey=1775.4$, not shown). In the opposite direction,
the wave also starts shrinking but the process stops, slightly
reverses and track of the wave is lost, as pointed out by the
diverging dotted trajectory. This trajectory wanders around a ghost of
the disappeared modulated wave for some time, then spirals away until
it comes close to the lower-branch travelling wave (plus signs) and
leaves the region towards the laminar flow. Close analysis seems to
indicate that the branch bends and turns around a fold (saddle-node)
of cycles picking up an unstable direction. The phase map picture is,
in this sense, incomplete. For each of the stable cycles (nodal
branch) there must be an unstable cycle (saddle branch). As $\Rey$
increases from the fold-of-cycles point, the nodal branch approaches
the upper branch of travelling waves until eventual collision at the
Hopf point, while the saddle branch grows larger and larger. It will
be argued later that, unless some exotic symmetry-breaking event takes
place, the unstable modulated wave must end up colliding with the
lower branch of travelling waves in a saddle-loop bifurcation, where
it ceases to exist without any other consequence.

As already mentioned, decreasing $\Rey$ causes both branches to merge
in a fold-of-cycles, leaving unstable travelling waves as the only
existing solutions. Any trajectory issued from the vicinity of the
upper-branch wave spirals away, temporarily orbits around the ghost of
the disappeared modulated wave, and then continues spiraling away
until the trajectory manages to go around the lower-branch travelling
wave and finally leaves towards the basic laminar profile
$(\varepsilon_{\rm 3D},(\bnabla p)_z)=(0,1)$ along the only unstable
direction of the travelling wave.

\begin{figure}
  \begin{center}
    \begin{tabular}{cccc}
      \raisebox{0.33\linewidth}{(\textit{a})}\hspace{-0.6cm} &
      \raisebox{0.02\linewidth}{\includegraphics[height=0.29\linewidth,clip]{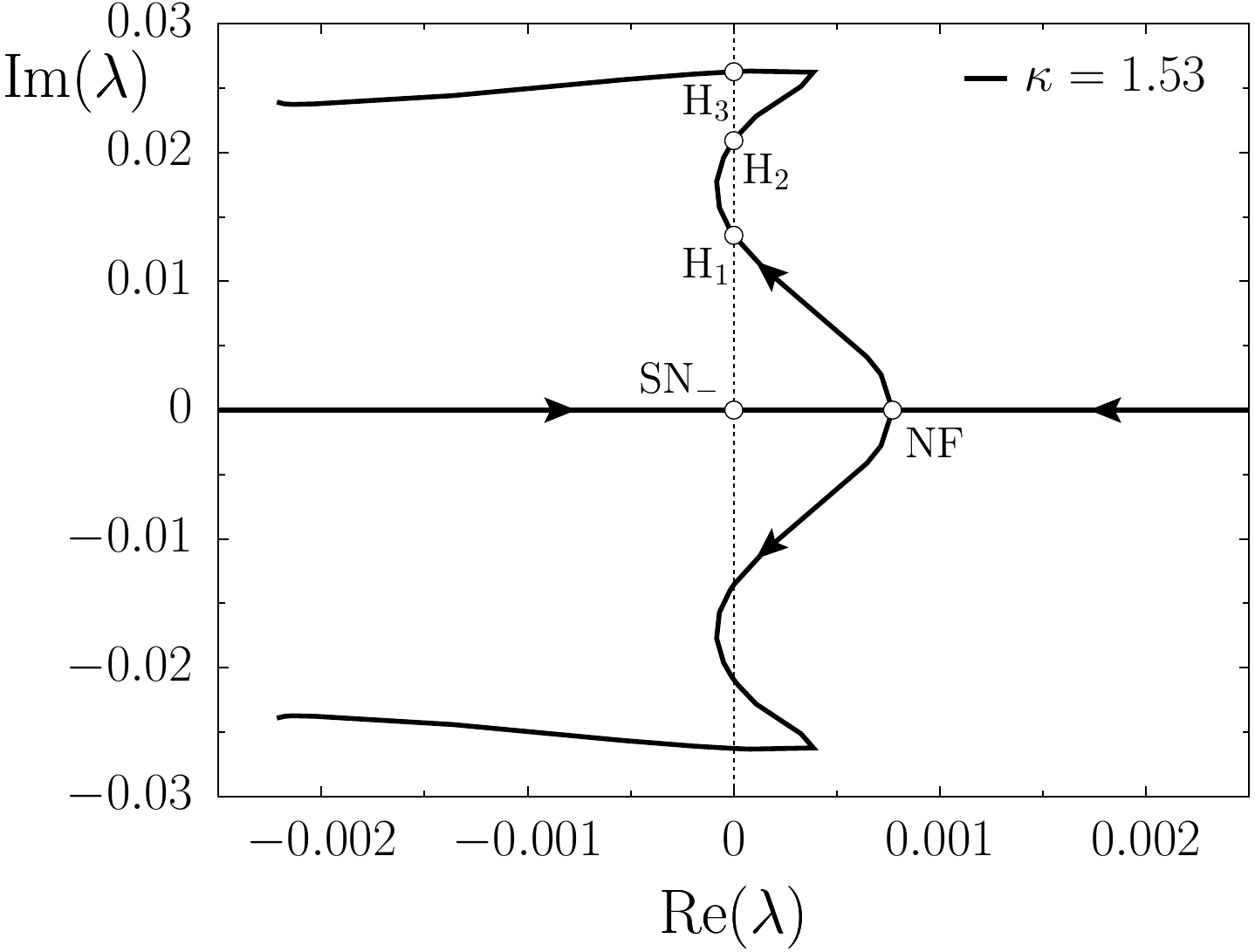}} &
      \raisebox{0.33\linewidth}{(\textit{b})}\hspace{-0.6cm} &
      \includegraphics[height=0.32\linewidth,clip]{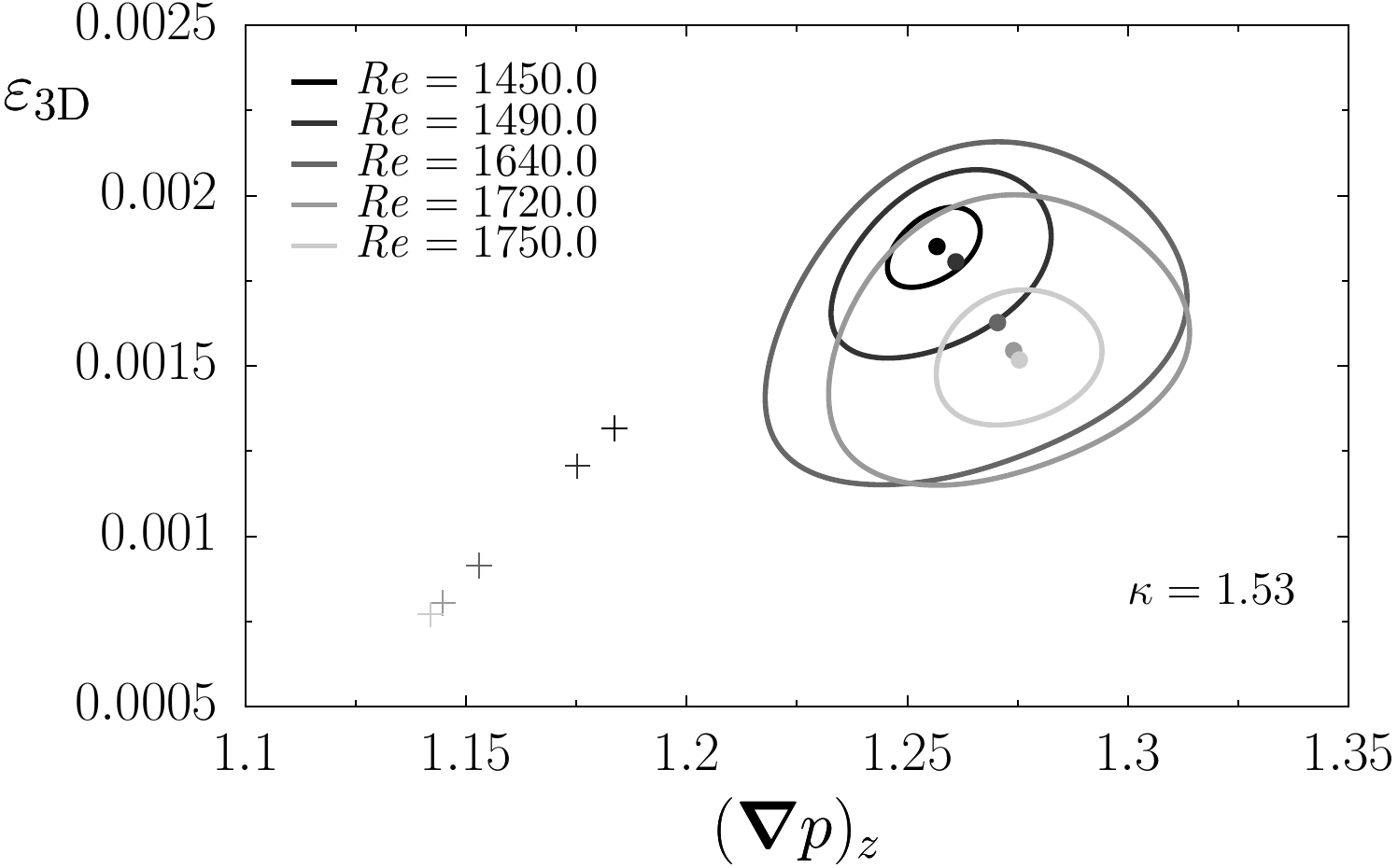}
    \end{tabular}
  \end{center}
  \caption{(\textit{a}) Eigenvalues trajectories for
    $\kappa=1.53$. (\textit{b}) Phase map projections of travelling
    and modulated waves along a $\Rey$-continuation for
    $\kappa=1.53$. See the caption of figure~\ref{fig:RPOPhMap} for
    explanations.}
  \label{fig:SpecPhMapk1.53}
\end{figure}

Continuation in the geometrical parameter $\kappa$ causes the
node-focus (${\rm NF}$) transition point to move around without
noticeable impact on the eigenvalues $\Rey$-trajectories, as can be
concluded from figures~\ref{fig:Spectra}(\textit{c},\textit{d}) and
\ref{fig:EV1.52}(\textit{a}). It becomes clear that for wavenumber
slightly beyond $\kappa \gtrsim 1.52$ the trajectories will experience
an additional couple of zero-crossings, adding a pair of Hopf
bifurcation points to the already existing one. This is illustrated
for $\kappa=1.53$ in figure~\ref{fig:SpecPhMapk1.53}(\textit{a}),
where the successive crossings along the upper branch have been
labelled ${\rm H_1}$, ${\rm H_2}$ and ${\rm H_3}$ as $\Rey$ is
increased.
Continuation of the modulated waves branch for $\kappa=1.53$, shown in
figure~\ref{fig:SpecPhMapk1.53}(\textit{b}), reveals that they
connect supercritically to the upper branch of travelling waves in
both directions at the two Hopf points that are at a larger distance
from the saddle-node bifurcation, \ie ${\rm H_2}$ and ${\rm H_3}$ of
figure~\ref{fig:SpecPhMapk1.53}(\textit{a}). The fold-of-cycles that
was present at $\kappa=1.52$ is no longer present at $\kappa=1.53$.

We shall extend this notation, ${\rm H_1}$, ${\rm H_2}$ and ${\rm
H_3}$, to denote the three potential Hopf crossings for any $\kappa$
in the range studied. It is clear that for decreasing $\kappa$, ${\rm
H_1}$ and ${\rm H_2}$ approach, collide and disappear leaving ${\rm
H_3}$ alone. Instead, for increasing $\kappa$ it is ${\rm H_2}$ and
${\rm H_3}$ that approach and collide, leaving ${\rm H_1}$ alone as
long as the node-focus point remains on the right side of the complex
plane.

\begin{figure}
  \begin{center}
      \includegraphics[height=0.45\linewidth,clip]{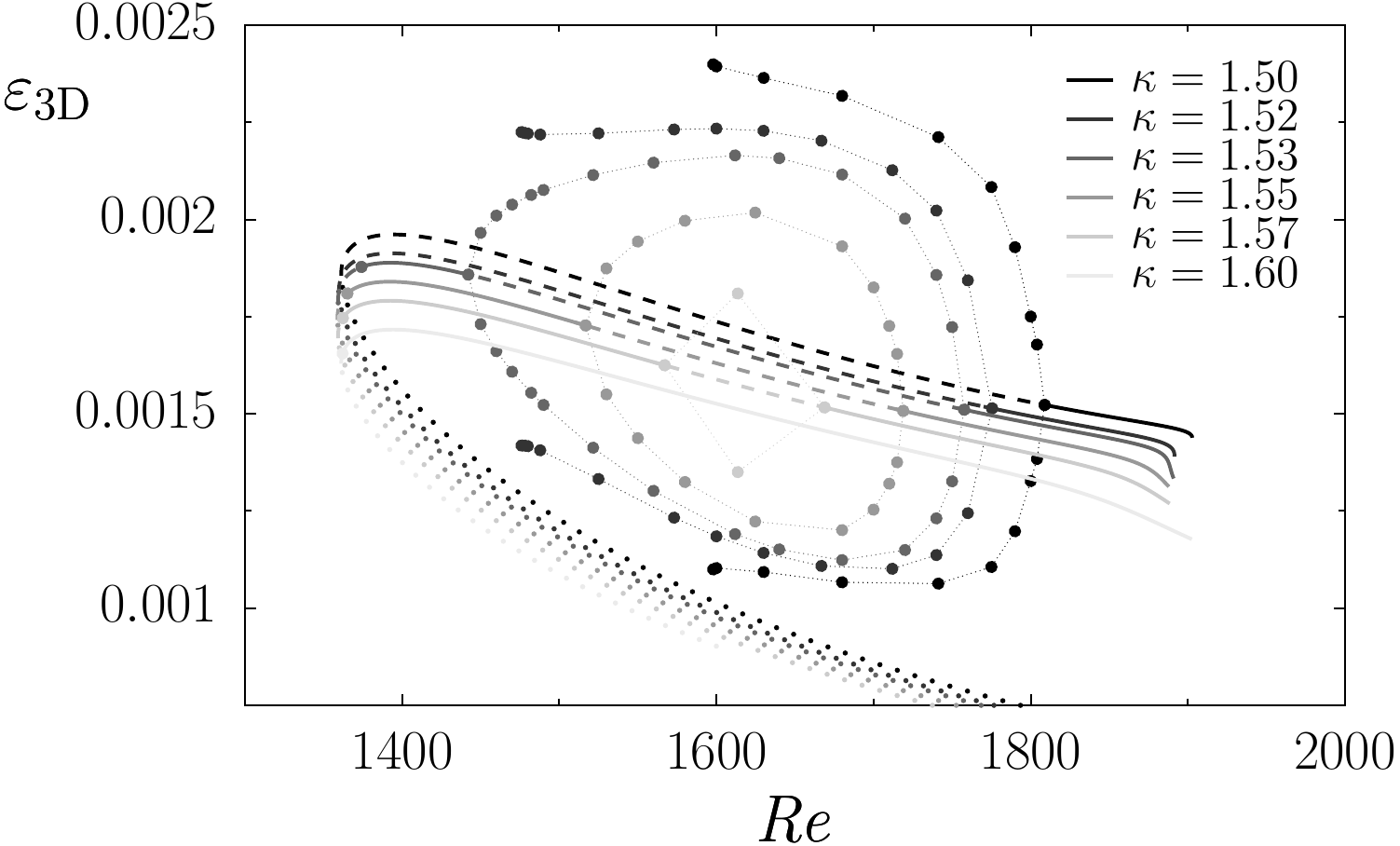}
  \end{center}
  \caption{Non-axisymmetric streamwise-dependent energy
    ($\varepsilon_{\rm 3D}$) as a function of $\Rey$ and $\kappa$
    (different gray intensities as indicated in the legend). Lines
    represent travelling waves: solid, dotted and dashed for $0$, $1$
    and $2$ unstable directions. Filled circles on the curves
    represent Hopf points. The rest of circles represent stable
    modulated waves.}
  \label{fig:EvsRek}
\end{figure}

Adding the $\Rey$-dependence helps to clarify the situation. This is
done in figure~\ref{fig:EvsRek}, where $\varepsilon_{\rm 3D}$ has been
plotted as a function of $\Rey$ and $\kappa$ (varying gray intensity
as indicated by the legend) for both travelling waves (solid, dotted
and dashed thick lines for $0$, $1$ and $2$-dimensional unstable
manifolds) and modulated waves (filled circles indicating minimum and
maximum values of $\varepsilon_{\rm 3D}$ along a period).
Lower-branch travelling waves have a single unstable direction. For
all cases shown, as they turn around the saddle-node point, a second
unstable direction appears. Low-$\kappa$ (\eg $\kappa=1.52$) waves
remain unstable until they regain stability at Hopf bifurcations,
whence supercritical branches of stable modulated waves are issued, as
already discussed for figure~\ref{fig:RPOPhMap}. Traces of the turning
point that we identified with a fold-of-cycles are clear from the
bending tips of the modulated wave branches. High-$\kappa$ (\eg
$\kappa=1.60$) waves regain stability almost immediately at Hopf
bifurcations but no stable branch of waves could be
detected. Intermediate-$\kappa$ (\eg $\kappa=1.53$) waves exhibit a
mixed behaviour by regaining stability early on the upper branch, but
losing it again temporarily for some range of $\Rey$. The whole
intervening unstable range is nevertheless {\em covered} by a stable
branch of modulated waves connecting both Hopf points, as illustrated
in figure~\ref{fig:SpecPhMapk1.53}(\textit{b}). At even higher
$\kappa$ (not shown), travelling waves become stable at the
saddle-node, as was the case for $\kappa=1.70$ of
figures~\ref{fig:Spectra}(\textit{b},\textit{d}). 
Returning to upper-branch travelling waves that were unstable at
onset, the branch of modulated waves that must be born at their early
stabilisation could not be detected with time evolution. This is a
clear sign that they may bifurcate subcritically, with the branch
pointing in the increasing $\Rey$ direction.

\begin{figure}
  \begin{center}
    \includegraphics[height=0.45\linewidth,clip]{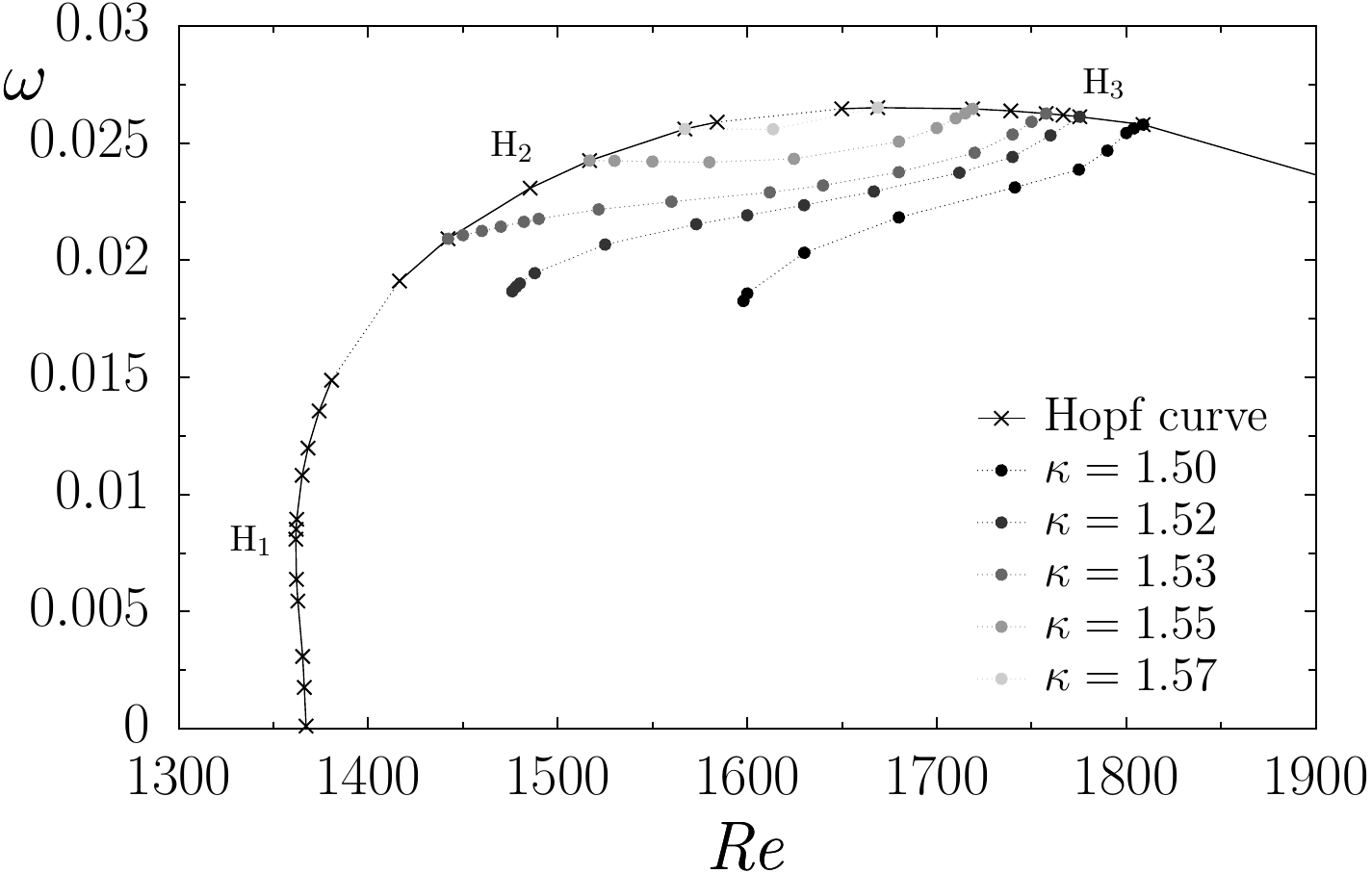}
  \end{center}
  \caption{Angular frequency ($\omega$) of the modulated waves and the
    Hopf points (filled circles). The thin line with crosses
    represents the imaginary part of the bifurcating eigenvalues along
    the locus of Hopf bifurcation points. The three segments, labelled
    ${\rm H_1}$, ${\rm H_2}$ and ${\rm H_3}$, correspond to
    intersections of the Hopf curve with iso-$\kappa$ lines.}
  \label{fig:fvsRek}
\end{figure}

Figure~\ref{fig:fvsRek} shows the angular frequency $\omega$ of the
modulated waves as a function of $\Rey$ and $\kappa$. Only a subset of
$\kappa$ has been plotted for clarity. And, of course, only as long as
the branch is stable, \ie accessible through time evolution.
Along with the waves, the locus of Hopf bifurcation points (thin
dotted line with crosses) has been added by plotting the imaginary
part of the bifurcating eigenvalues. Along this curve, $\kappa$ varies
continuously but not monotonically. The curve has been split in three
monotonical segments corresponding to the up to three independent
zero-crossings that the complex pair of eigenvalues might undergo
(${\rm H_1}$, ${\rm H_2}$ and ${\rm H_3}$). Naturally, the
modulational frequency of the bifurcating modulated waves is dictated
by the imaginary part of the eigenvalues at the Hopf bifurcation. This
figure strongly endorses the fold-of-cycles hypothesis for the
non-reconnecting branches. The decreasing frequency around the
fold-of-cycles is also suggestive of the eventual saddle-loop
collision with lower-branch travelling waves, which demands a
vanishing frequency (period going to infinity) as the modulated wave
transforms into a non-robust homoclinic loop and disappears.

It is clear that for $1.52 \lesssim \kappa \lesssim 1.53$ a change of
behaviour takes place. Modulated wave branches go from reconnecting to
the Hopf curve to bending and losing stability. It seems therefore
reasonable that for some intermediate value, a branch of solutions may
become tangent to the Hopf curve, precisely at the $\kappa$ value for
which the couple of additional eigenvalue crossings start occurring,
\ie where ${\rm H_1}$ and ${\rm H_2}$ merge. Slightly above this
value, the curve would split in two. One of the segments connects
supercritically two of the three existing Hopf points (${\rm H_2}$ and
${\rm H_3}$). The other segment would be issued from the third Hopf
point (${\rm H_1}$) and, for purely geometrical reasons, it would have
to do so supercritically towards lower $\Rey$. Unfortunately, the room
left in parameter space for 
the stable branch of waves before it bends back and becomes unstable
is tiny and very close to the Hopf bifurcation line where the relevant
complex pair of eigenvalues exhibits a vanishing real part. As a
consequence, transients are far too long to allow computation of the
stable modulated waves in a feasible time-scale. This ultimately
renders the numerical determination of the borders of existence of
such solutions within this region extremely difficult. For even higher
$\kappa$, but not exceeding the value for which upper-branch
travelling waves are stable at onset, modulated waves can only
bifurcate subcritically from ${\rm H_1}$ towards increasing $\Rey$ and
with vanishing frequencies, as required by the double-zero bifurcation
scenario that takes place.


\subsection{Unfolding of the Takens-Bogdanov bifurcation}
\label{subsec:takbog}

\begin{figure}
  \begin{center}
    \includegraphics[width=0.8\linewidth,clip]{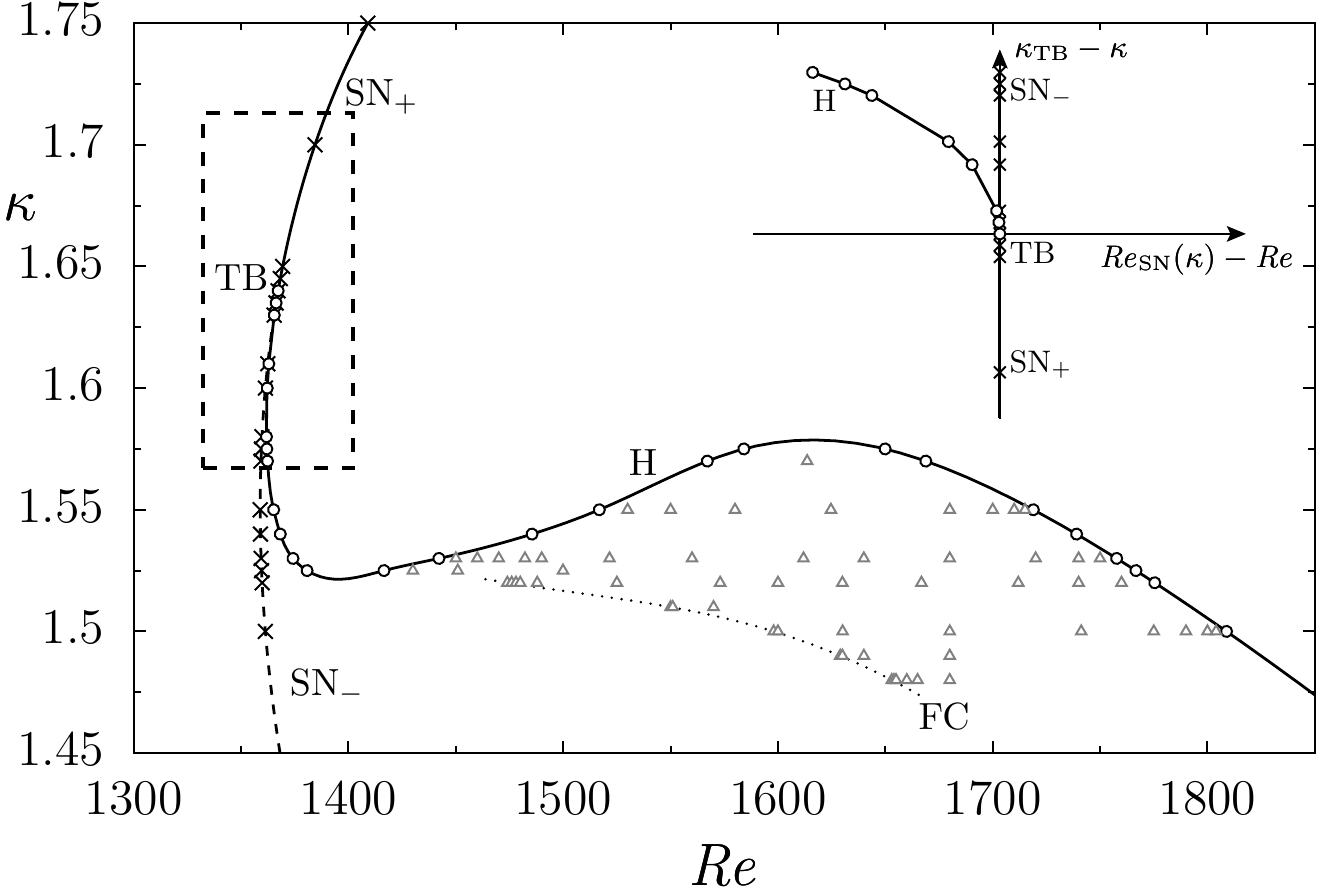}
  \end{center}
  \caption{Bifurcation curves in $(\kappa,\Rey)$-parameter
  space. ${\rm SN_+}$ (solid line with crosses) and ${\rm SN_-}$
  (dashed line with crosses) represent stable and unstable saddle-node
  bifurcations of travelling waves. ${\rm H}$ (solid line with
  circles) is the Hopf bifurcation curve of upper-branch travelling
  waves. ${\rm TB}$ is the Takens-Bogdanov (double-zero) bifurcation
  point. Stable modulated waves are indicated with triangles, with
  ${\rm FC}$ (dotted line) representing a fold-of-cycles. The inset
  shows a transformed zoom of the dashed region, with new coordinate
  axes $\kappa_{\rm TB}-\kappa$ and $\Rey_{\rm SN}(\kappa)-\Rey$.}
  \label{fig:kvsRe}
\end{figure}

The $2$-fold azimuthally-periodic, shift-reflect symmetric family of
travelling waves has been continued both in $\Rey$ and $\kappa$, and
systematic linear stability analysis has been carried out. In this
manner, the locus of the saddle-node bifurcation between upper- and
lower-branch travelling waves, and that of the Hopf bifurcation of
upper-branch travelling waves has been tracked in parameter
space. Results are summarised in figure~\ref{fig:kvsRe}.
To the left of the saddle-node bifurcation lines (${\rm SN_+}$ and
${\rm SN_-}$), there exist no travelling waves. These appear in
saddle-node bifurcations to the right of the saddle-node
lines. Upper-branch waves produced in ${\rm SN_+}$ are stable nodes,
whereas those appearing in ${\rm SN_-}$ are unstable, all lower-branch
travelling waves being unstable with a single unstable
direction. Almost immediately after springing into existence,
upper-branch travelling waves experience a node-focus transition
(${\rm NF}$ of figure~\ref{fig:Spectra}). This non-bifurcation curve
is not shown in the figure, as it is extremely close to the
saddle-node curve.  ${\rm SN_+}$, ${\rm SN_-}$ and ${\rm NF}$ emanate
from a Takens-Bogdanov or double-zero bifurcation point (${\rm
TB}$). At this point, two real eigenvalues collide exactly at the
origin of the complex plane and become a conjugate pair. From this
very same point, a Hopf bifurcation line (${\rm H}$) is issued
downwards to the right of ${\rm SN_-}$. Initially, the lower is
$\kappa$, the further away ${\rm H}$ gets from ${\rm SN_-}$. This is a
consequence of the left-opening parabolic shape of the complex
eigenvalues (figure~\ref{fig:Spectra}). As $\kappa$ is reduced, ${\rm
NF}$ moves away from the origin and the distance along the trajectory,
which is related to $\Rey$, before crossing gets longer. Further along
${\rm H}$, a couple of successive upward and downward bendings
occur. The three segments into which they split the curve can be
identified with ${\rm H_1}$, ${\rm H_2}$ and ${\rm H_3}$, the three
potential crossings the eigenvalues may undergo at any given
$\kappa$. Above this line, upper-branch travelling waves are
stable. Below it they are always unstable. As a consequence of its
last downwards turn, for sufficiently large $\Rey$ all upper-branch
solutions are left stable as far as this bifurcation diagram is
concerned and no secondary instabilities take place. To complete the
phenomenological description, modulated waves have also been added
into the picture (empty triangles). They occupy a region enclosed
between the Hopf line and another line defined as the locus of turning
points where modulated waves cease to exist. This line is a
bifurcation curve corresponding to a fold-of-cycles (${\rm FC}$),
whereby stable modulated waves turn around and become unstable.

\section{Discussion}\label{sec:discu}


It is at this point convenient to define an alternative couple of
parameters: $\kappa_{\rm{TB}}-\kappa$ and $\Rey_{\rm
SN}(\kappa)-\Rey$, where $\kappa_{\rm{TB}}$ is the value of $\kappa$
at the Takens-Bogdanov point and $\Rey_{\rm SN}(\kappa)$ is the value
of $\Rey$ at the saddle-node point for each given $\kappa$. The
transformation has been applied to the region indicated by the dashed
square in figure~\ref{fig:kvsRe} and plotted in the inset. The
resulting bifurcation diagram is reminiscent of the double-zero
bifurcation described in \cite*{Wiggins_B_03}. Of the two equivalent
normal forms proposed for this bifurcation by \cite{Takens_IHES_74}
and \cite{Bogdanov_FAA_75}, the former can be applied to the present
bifurcation by simply taking $\mu_1 \simeq \kappa_{\rm{TB}}-\kappa$
and $\mu2 \simeq \Rey_{\rm SN}(\kappa)-\Rey$ in \cite{Wiggins_B_03}:
\begin{equation}
  \left\{
  \begin{array}{l}
    \dot{x}=y,\\
    \dot{y}=\mu_1 + \mu_2 y + x^2 + b x y,
  \end{array} \right.
  \label{eq:WigginsTB}
\end{equation}
with $b=+1$. A detailed analysis of the alternative normal form can be
found in \cite*{Kuznetsov_B_95}.

This second-order normal form provides an accurate description of all
the phenomenology observed in the immediate vicinity of the
Takens-Bogdanov point and gives additional information not accessible
with the methodology here deployed. From the Takens-Bogdanov (${\rm
TB}$) point (inset of figure~\ref{fig:kvsRe}), stable (${\rm SN_+}$)
and unstable (${\rm SN_-}$) saddle-node lines extend along the
vertical axis downwards and upwards, respectively. Issued from the
same point is a subcritical Hopf curve (${\rm H}$) with a left-opening
half-parabolic shape, to the left of the unstable saddle-node
line. The unstable node created at the saddle-node immediately becomes
a focus and later recovers stability upon crossing the Hopf curve
leftwards. An unstable branch of unstable periodic orbits originates
surrounding the stabilised equilibrium. Close analysis of the
bifurcation reveals that the unstable periodic orbits grow large and
collide with the unstable saddle in a saddle-loop homoclinic
bifurcation, not shown in the figure. The unstable periodic orbits
disappear with no consequence for the saddle equilibrium, leaving it,
together with the stable focus, as the only surviving solutions. A
saddle-loop curve, also branching off the Takens-Bogdanov point, must
therefore be present, although it cannot be detected with the methods
available.


Analysis of the second-order normal form is nevertheless insufficient
to account for the existence of stable modulated waves and the back
and forth bending of the Hopf curve. It is however possible to analyse
in some detail what the effect of higher-order terms would be if the
conditions for a non-degenerate double-zero bifurcation were nearly
violated and higher-order terms were not negligible some distance away
from the bifurcation point. A complete analysis of the codimension-$3$
bifurcation resulting from a vanishing second-order coefficient ($b$
in \ref{eq:WigginsTB}) can be found in \cite{DuRoSoZo_B_91}. It is
unlikely, unless some symmetry at play cancels exactly some
second-order coefficients, which is not the case (the eigenfunctions
involved preserve all symmetries of the solution), that simple tuning
of two parameters will cross a codimension-$3$ point. Therefore, it is
not the codimension-$3$ point we are interested in, but a perturbation
around it that renders higher-order terms important at some distance
from the codimension-$2$ point under analysis.

Such an analysis was carried out by \cite*{Barkley_PoFA_90} when
studying similar phenomena in plane Poiseuille flow, \ie the
appearance of modulated waves in channel flow. Previous studies
\cite*[][]{PuSa_JFM_88,SoMe_JFM_91} had reported the onset of a
modulational instability of travelling waves when the flow was driven
by a constant pressure gradient. No instability was detected for the
same waves when the flow was driven at constant mass-flux. Introducing
a new parameter allowing to connect continuously both possible
boundary conditions, a bifurcating scenario arose as a natural
explanation. It was suggested that while a double-zero bifurcation
could explain the appearance of unstable modulated waves, the simple
inclusion of third-order terms to the second-order normal form could
explain the existence of stable modulated waves for some parameter
range. According to normal form theory, the extended normal form needs
only include a couple of third-order terms
\cite*[][]{Guckenheimer_B_83,Knobloch_PLA_86,IoosAdelmeyer_B_98}
\begin{equation}
  \left\{
  \begin{array}{l}
    \dot{x}=y,\\
    \dot{y}=\mu_1 + \mu_2 y + x^2 + k_1 x y + k_2 x^3 + k_3 x^2 y,
  \end{array} \right.
  \label{eq:WigginsTB3}
\end{equation}
where appropriate scaling can be performed such that $k_3=\pm 1$. The
second order coefficient $k_1$ is required to be sufficiently small so
that third-order terms come into play, but non-zero so that
non-degeneracy of the regular Takens-Bogdanov is preserved for
sufficiently small $\mu_1$ and $\mu_2$.

We shall not reproduce here the analysis of the third-order-extended
normal form for the Takens-Bogdanov bifurcation. Instead, we refer the
reader to \cite{Barkley_PoFA_90} for a detailed unfolding, and pick up
on the main results that are relevant to the present case. It can be
shown that appropriate tuning of the coefficients enables this normal
form to capture some of the phenomenology observed but so far
unexplained. Namely, setting $k_3=+1$ and $k_1>0$, gives the right
eigenvalues path up to the first right turn after the node-focus
transition (figures~\ref{fig:Spectra}\textit{c},\textit{d} and
\ref{fig:EV1.52}\textit{a}). As a consequence, the Hopf curve
experiences a first turning point as seen in figure~\ref{fig:kvsRe},
allowing for multiplicity of Hopf points at any given $\kappa$ above
some minimum value. A most relevant result concerns the nature of the
Hopf curve itself. It has already been mentioned that in the vicinity
of the Takens-Bogdanov point, the Hopf curve happens to be subcritical
in the sense that the first Lyapunov coefficient is positive and
bifurcating periodic orbits are unstable. Restricting
(\ref{eq:WigginsTB3}) to the Hopf curve and with an appropriate change
of variables and rescaling, we obtain
\begin{equation}
  \left\{
  \begin{array}{l}
    \dot{\rho}=\rho (\beta_1 + \beta_2 \rho^2 + s \rho^4) + O(\rho^7),\\
    \dot{\phi}=1,
  \end{array} \right.
  \label{eq:KuznetsovB}
\end{equation}
with $s=+1$, which is the normal form for the degenerate Hopf or
Bautin bifurcation \cite[][]{Kuznetsov_B_95}. $\beta_2 \sim k_1-3 k_2
\mu_2$ is the first Lyapunov coefficient and governs the change in
criticality along the Hopf curve. Taking $k_2>0$ will have $\beta_2$
vanish for some value of $\mu_2$, leaving the Hopf curve supercritical
beyond some point sufficiently far from the Takens-Bogdanov
point. This is exactly what is needed to account for supercritical
bifurcation of stable modulated waves. Careful analysis of the Bautin
bifurcation also requires that the stable modulated waves turn around
in a fold-of-cycles (${\rm FC}$), whenever they cannot branch to
another Hopf point, becoming unstable. Since the original unstable
waves collided with the saddle in a saddle-loop bifurcation in the
frame of the Takens-Bogdanov bifurcation, it is to be expected that
they will continue exhibiting the same behaviour even if they now
initially bifurcate supercritically. Thus, even though we could only
follow stable modulated waves to their fold bifurcation
(figure~\ref{fig:EvsRek}), we can now reliably comment on their fate:
they should most certainly disappear in saddle-loop bifurcations. This
was already justified upon explaining figure~\ref{fig:fvsRek}, where
the trend of the angular frequency suggested that the period could go
to infinity after turning around, as the saddle-loop bifurcation
requires for the homoclinic connection in which they convert before
disappearing. The Bautin point must be somewhere along the Hopf curve
before the first turning point. As it happens, supercriticality is
only extremely mild until about the turning point (the locus of the
fold-of-cycles appears to be very close to the Hopf curve), resulting
in our inability to compute the stable modulated waves that must be
present. This is so because close to the Hopf curve, the dynamics are
extremely slow and the transients accordingly long. These are the
waves that are expected to turn around, become unstable and are
destroyed in saddle-loops as predicted by third-order terms. Beyond
the turning point, supercriticality means branching towards lower
$\mu_1$ (higher $\Rey$). These waves we have computed, but their fate
crucially depends upon the second turning of the Hopf curve, and
third-order terms tell us nothing about it.

Adding fourth-order terms is a straightforward but tedious calculation
that we shall not reproduce here. It cannot be expected that
fourth-order amplitude equations capture the behaviour reliably so far
from the Takens-Bogdanov point, yet there is strong numerical evidence
that fourth-order terms have a say. They allow for an extra turning
point to the Hopf curve, without altering any of the previous
predictions. As a result, stable modulated waves born between the two
turning points can disappear in a second supercritical Hopf after the
second turn, as illustrated in figures~\ref{fig:EvsRek} and
\ref{fig:kvsRe}. Further down the Hopf curve, beyond the second
turning point, stable modulated waves either behave us just explained
or progress stably until they meet the fold-of-cycles (${\rm FC}$) and
turn around. This curve must 
emerge from the Hopf curve precisely at
the Bautin point, the existence of which can be inferred but its
location not determined precisely with the tools at hand.

\begin{figure}
  \begin{center}
    \includegraphics[width=0.8\linewidth,clip]{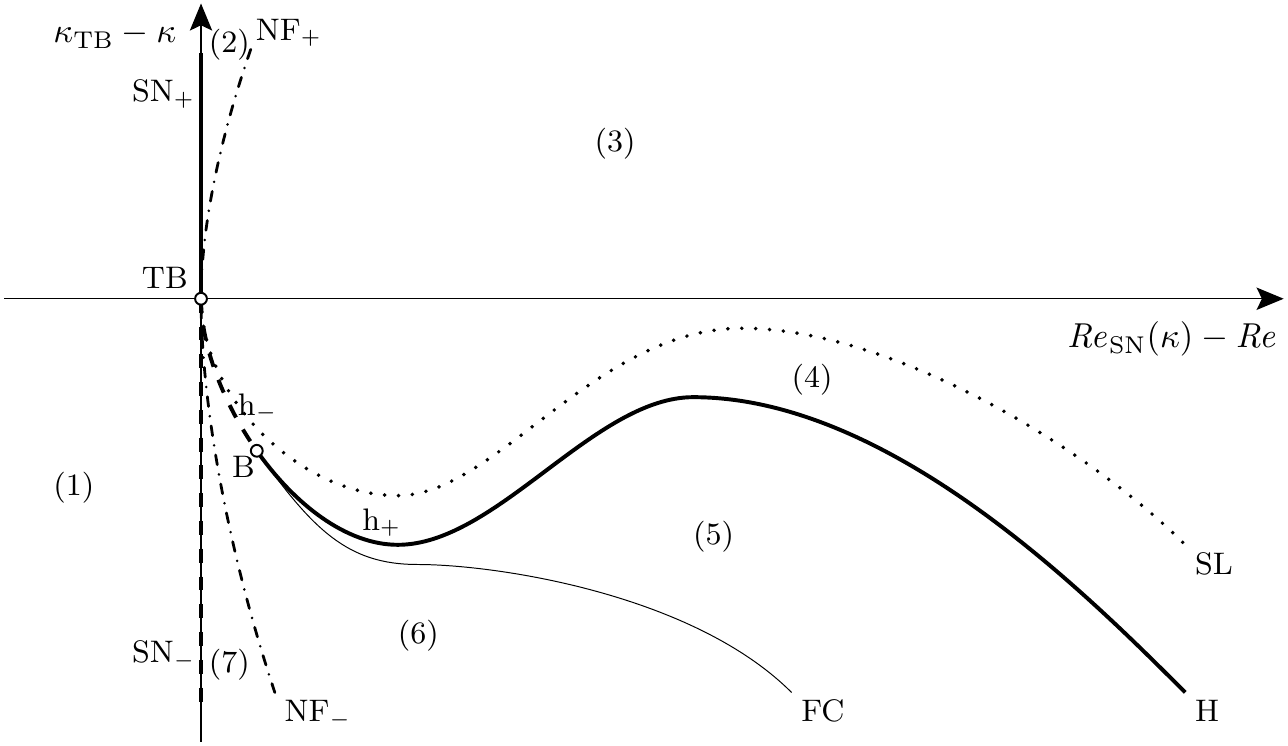}
  \end{center}
  \caption{Attemptive bifurcation diagram in $(\kappa_{\rm
  TB}-\kappa,\Rey_{\rm SN}(\kappa)-\Rey)$-parameter space. Represented
  are the two codimension-$2$ points ${\rm TB}$ (double zero or
  Takens-Bogdanov) and ${\rm B}$ (Bautin or degenerate Hopf). ${\rm
  B}$ splits the Hopf curve (${\rm H}$) in two segments, one
  subcritical (${\rm h_-}$, dashed thick line), the other
  supercritical (${\rm h_+}$, solid thick line). From ${\rm TB}$,
  stable and unstable saddle-node curves (${\rm SN_+}$ and ${\rm
  SN_-}$), the Hopf curve and a saddle-loop curve (${\rm SL}$, dotted
  line) are born, together with stable and unstable node-focus
  transition curves (${\rm NF_+}$ and ${\rm NF_-}$, dash-dotted
  lines). A fold of cycles (${\rm FC}$) is issued from ${\rm
  B}$. Numbers in parenthesis indicate different regions.}
  \label{fig:TBbif}
\end{figure}

Figure~\ref{fig:TBbif} completes the picture just sketched in an
endeavouring diagram for all bifurcations in parameter space. Modified
parameters $(\kappa_{\rm TB}-\kappa,\Rey_{\rm SN}(\kappa)-\Rey)$ have
been chosen for simplicity.
The full diagram is organised around the Takens-Bogdanov (${\rm TB}$)
codimension-$2$ point. This point separates two branches of
saddle-node lines (${\rm SN_+}$ and ${\rm SN_-}$). At ${\rm SN_+}$ a
stable node and a saddle equilibrium coalesce, while crossing ${\rm
  SN_-}$ generates an unstable node and a saddle. The node, stable or
unstable, experiences a quick transition and becomes a focus, stable
or unstable, in a node-focus transition (${\rm NF_+}$ and ${\rm
  NF_-}$). These are non-bifurcation curves. From ${\rm TB}$, a Hopf
bifurcation curve ${\rm H}$ springs quadratically to the same side as
${\rm SN_-}$. Sufficiently close to ${\rm TB}$ periodic orbits
bifurcate subcritically from the node at ${\rm H}$ (${\rm
  h_-}$). Further away, ${\rm H}$ changes criticality at a
codimension-$2$ Bautin point (${\rm B}$, degenerate Hopf) and becomes
supercritical (${\rm h_+}$). At ${\rm B}$, a fold-of-cycles curve
(${\rm FC}$) is born to the ${\rm h_+}$ side of ${\rm H}$ and parting
from it quadratically. Between ${\rm h_+}$ and ${\rm FC}$, stable
periodic orbits exist. At ${\rm h_-}$ and ${\rm FC}$, unstable
periodic orbits are generated. They must all cease to exist, at least
in the vicinity of ${\rm TB}$, in saddle-loop bifurcations (${\rm
  SL}$). At ${\rm SL}$, the unstable periodic orbit has grown large
and collides with the saddle producing a non-robust homoclinic
cycle. The periodic orbit vanishes while the saddle experiences no
change whatsoever. Among all bifurcation curves drawn in
figure~\ref{fig:TBbif}, ${\rm SL}$ is the most uncertain. Since it
runs on the saddle, independent from the nodal equilibrium, there is
no restriction on whether this line can cross ${\rm H}$. Thorough
analysis of the Takens-Bogdanov bifurcation shows that it cannot cross
${\rm h_-}$ and that it actually departs quadratically from ${\rm
  SN_-}$, but higher-order terms can easily make it cross ${\rm
  h_+}$. Unfortunately no linear stability analysis of the saddle can
cast light on the issue, nor can the unstable periodic orbits be
tracked with time evolution. However, the picture just sketched would
not change much.

\begin{figure}
  \begin{center}
    \includegraphics[width=0.9\linewidth,clip]{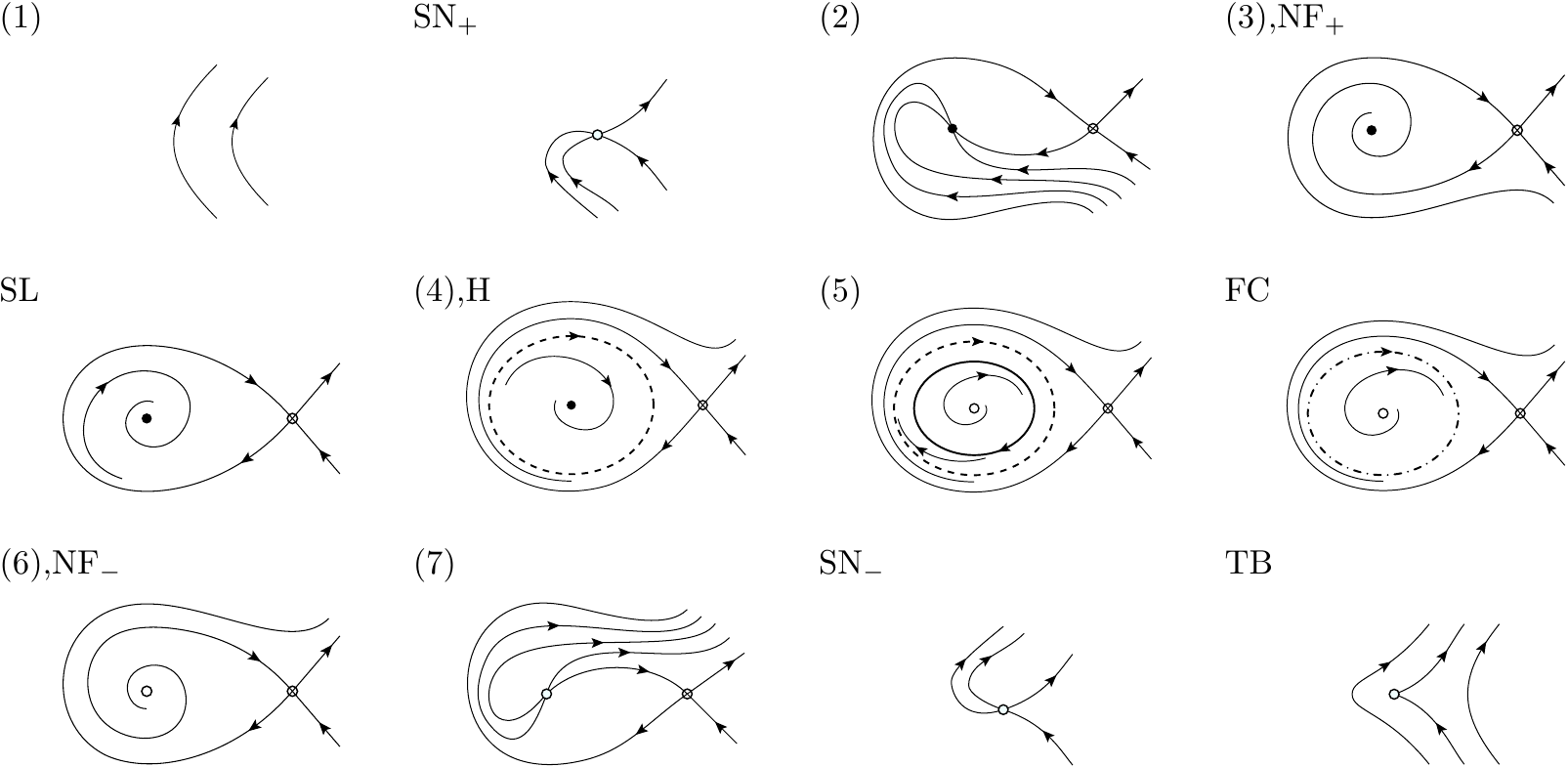}
  \end{center}
  \caption{Phase map diagrams corresponding to all bifurcation curves
  and numbered regions in figure~\ref{fig:TBbif}. Small circles
  correspond to relative equilibria (filled for stable equilibria),
  and closed thick lines to relative periodic orbits (solid for
  stable, dashed for unstable orbits). Thin lines with arrows indicate
  stable and unstable manifolds, homoclinic and heteroclinic
  connections.}
  \label{fig:PhMaps}
\end{figure}

Figure~\ref{fig:PhMaps} shows phase map diagrams of the flow projected
on the 2-dimensional centre manifold. The labels indicate to which
region in figure~\ref{fig:TBbif} each phase map corresponds.
Making a clockwise round-trip can help clarify the picture. Starting
from $(1)$, where no solutions exist, and crossing ${\rm SN_+}$ into
$(2)$ a saddle and a stable node are created. The stable node turns
into a stable focus through ${\rm NF_+}$ in $(3)$. In $(4)$, an
unstable periodic orbit exists, that has been created in ${\rm
SL}$. Upon crossing ${\rm h_+}$ into $(5)$ a stable cycle appears
around the focus, which becomes unstable. The coexistence of an
unstable and a stable cycle in $(5)$ lasts until they collide and
disappear in ${\rm FC}$, leaving only the saddle and an unstable focus
in $(6)$. It is possible to arrive to $(6)$ directly from $(4)$. In
this case, the unstable cycle just merges with the stable focus in
${\rm h_-}$ to leave an unstable focus. Crossing ${\rm NF_-}$ the
unstable focus transforms into an unstable node in $(7)$. The unstable
focus coalesces with the saddle through ${\rm SN_-}$ back into region
$(1)$. The possible crossing of ${\rm SL}$ and ${\rm h_+}$ would only
affect the order in which the stable cycle merges with the focus and
the unstable cycle collides with the saddle, when going from $(5)$ to
$(3)$.


It is at this point necessary to justify why we have not made use of
the discrete azimuthal periodicity or of the shift-reflect symmetry in
deriving the normal form. The fact is that we are entitled to ignore
these symmetries because they play no significant role in the problem
at hand. If any of the solutions had broken any of the symmetries,
these would have needed to be built into the normal form. Since there
is no numerical evidence from either time evolution or linear
stability that this happens, these symmetries can be safely ignored.

\section{Conclusion}\label{sec:conclu}


A complex bifurcating scenario in pipe Poiseuille flow, involving the
appearance of travelling as well as modulated travelling waves, has
been thoroughly analysed within a given azimuthal subspace, taking
advantage of the special stability properties of the solutions it is
home to. These stability properties make this subspace particularly
fit for the search and analysis of all kinds of exotic behaviour
including, but not limited to, the double-zero bifurcation here
unfolded. This can help place a lower bound to the extreme complexity
that must certainly be present in full space pipe flow.




The particular family of waves studied involves the appearance in Hopf
bifurcations of modulated waves that are either unstable at onset or
become unstable in a fold-of-cycles. It cannot 
be guaranteed that the unstable modulated waves created in the
fold-of-cycles eventually disappear in a saddle-loop
bifurcation. Their ultimate fate depends on the absence of
symmetry-breaking bifurcations that free them from the need of
colliding with the saddle or that allow for an entanglement of the
stable and unstable manifolds of the saddle in the vicinity of the
global bifurcation. 
While such a behaviour has to be expected generically, the absence of
chaotic dynamics around the saddle suggests that this is not the case,
and that the scenario described may be fairly accurate.


The Takens-Bogdanov bifurcation experienced by this particular family
of travelling waves must not be considered as an exotic
scenario. There is strong evidence that other families of travelling
waves might be bifurcating in similar environments. This is the case
of the highly-symmetric $3$-fold azimuthally-periodic family of
travelling waves described in \cite{PrDuKe_PTRSA_09}, for which the
eigenvalue associated with the saddle-node bifurcation is known to
collide with another eigenvalue and go complex immediately after
crossing from lower to upper branch. This happens in the context of
instability dictated by several additional unstable eigenvalues, which
renders the techniques used in the present study unable to unfold the
bifurcation. Nevertheless, a Takens-Bogdanov bifurcation appears to be
there all the same.


The scenario is nonetheless not general. Other families of travelling
waves, such as the mirror-symmetric $2$-fold azimuthally-periodic, do
not follow this behaviour. Other travelling waves come from Pitchfork
bifurcations from other more symmetric waves, \eg the shift-reflect
azimuthally non-periodic travelling wave \cite[][]{PrKe_PRL_07}, which
comes from symmetry breaking of a mirror-symmetric wave. The
Takens-Bogdanov bifurcation simply does not apply in these cases.


As a result of a Takens-Bogdanov bifurcation, no trace of modulated
waves is expected to survive long enough to take part in the formation
of a chaotic attractor or saddle as a precursor to turbulent
dynamics. Nevertheless, it is interesting to point out that, far
enough from the Takens-Bogdanov point, only stable upper-branch
solutions are left. Thus, all the techniques here deployed are left
valid for an eventual analysis of the bifurcation cascade they may
undergo at higher $\Rey$. A detailed study of the behaviour of these
waves at higher flow rates is currently underway.



\begin{acknowledgments}
This work has been supported by the Spanish Ministry of Science and
Technology, under grants FIS2007-61585, by the Catalan
Government under grant SGR-00024, and by the Deutsche
Forschungsgemeinschaft.

\end{acknowledgments}


\bibliographystyle{jfm}
\bibliography{JFM09}

\begin{thebibliography}{57}
\expandafter\ifx\csname natexlab\endcsname\relax\def\natexlab#1{#1}\fi

\bibitem[Barkley({1990})]{Barkley_PoFA_90}
{\sc Barkley, D.} {1990} {Theory and predictions for finite-amplitude waves in
  2-dimensional plane Poiseuille flow}. {\em Phys.\ Fluids\,A\/} {\bf
  {2}}~({6}), {955--970}.

\bibitem[Boberg \& Brosa(1988)]{BoBr88}
{\sc Boberg, L. \& Brosa, U.} 1988 Onset of turbulence in a pipe. {\em Z.
  Naturforsch. A: Phys. Sci.\/} {\bf 43}, 697--726.

\bibitem[Bogdanov({1975})]{Bogdanov_FAA_75}
{\sc Bogdanov, R.} {1975} {Versal deformations of a singular point on the plane
  in the case of zero eigenvalues}. {\em {Func.\ Anal.\ Appl.}\/} {\bf
  {9}}~({144--145}).

\bibitem[Brosa \& Grossmann(1999)]{BrGr99}
{\sc Brosa, U. \& Grossmann, S.} 1999 Minimum description of the onset of pipe
  turbulence. {\em Eur.\ Phys.\ J.\,B\/} {\bf 9}~(2), 343--354.

\bibitem[Chossat \& Lauterbach(2000)]{ChossatLauterbach_B_00}
{\sc Chossat, P. \& Lauterbach, R.} 2000 {\em Methods in Equivariant
  Bifurcations and Dynamical Systems\/}. London: World Scientific Publishing.

\bibitem[Darbyshire \& Mullin(1995)]{DaMu_JFM_95}
{\sc Darbyshire, A.G. \& Mullin, T.} 1995 Transition to turbulence in
  constant-mass-flux pipe flow. {\em J.\,Fluid Mech.\/} {\bf 289}, 83--114.

\bibitem[Dennis \& Schnabel(1996)]{DeSc_B_96}
{\sc Dennis, J.E. \& Schnabel, R.B.} 1996 {\em Numerical Methods for
  Unconstrained Optimization and Nonlinear Equations\/}. Englewood Cliffs,
  N.J.: SIAM.

\bibitem[Duguet {\em et~al.\/}(2008)Duguet, Willis \& Kerswell]{DuWiKe_JFM_08}
{\sc Duguet, Y., Willis, A.P. \& Kerswell, R.R.} 2008 Transition in pipe flow:
  the saddle structure on the boundary of turbulence. {\em J.\,Fluid Mech.\/}
  {\bf 613}, 255--274.

\bibitem[Dumortier {\em et~al.\/}(1991)Dumortier, Roussarie, Sotomayor \&
  Zoladek]{DuRoSoZo_B_91}
{\sc Dumortier, F., Roussarie, R., Sotomayor, J. \& Zoladek, H.} 1991 {\em
  Bifurcations of Planar Vector Fields. Nilpotent Singularities and Abelian
  Integrals\/}. Berlin: Springer-Verlag.

\bibitem[Eckhardt(2009)]{Eckhardt09}
{\sc Eckhardt, B.} 2009 {Introduction. Turbulence transition in pipe flow:
  125th anniversary of the publication of Reynolds' paper}. {\em Phil.\ Trans.\
  Roy.\ Soc.\ Lond.\,A\/} {\bf 367}~(1888), 449--455.

\bibitem[Eckhardt {\em et~al.\/}(2007)Eckhardt, Schneider, Hof \&
  Westerweel]{ESHW91}
{\sc Eckhardt, B., Schneider, T.M., Hof, B. \& Westerweel, J.} 2007 Turbulence
  transition in pipe flow. {\em Ann.\ Rev.\ Fluid Mech.\/} {\bf 39}, 447--468.

\bibitem[Ehrenstein \& Koch({1991})]{EhKo_JFM_91}
{\sc Ehrenstein, U. \& Koch, W.} {1991} {3-dimensional wave-like equilibrium
  states in plane Poiseuille flow}. {\em J.\,Fluid Mech.\/} {\bf {228}},
  {111--148}.

\bibitem[Faisst \& Eckhardt(2003)]{FaEc_PRL_03}
{\sc Faisst, H. \& Eckhardt, B.} 2003 Travelling waves in pipe flow. {\em
  Phys.\ Rev.\ Lett.\/} {\bf 91}~(22), 224502.

\bibitem[Frayss\'e {\em et~al.\/}(2003)Frayss\'e, Giraud, Gratton \&
  Langou]{CERFACS_03}
{\sc Frayss\'e, V., Giraud, L., Gratton, S. \& Langou, J.} 2003 A set of gmres
  routines for real and complex arithmetics on high performance computers.
  Technical Report TR/PA/03/3. CERFACS, http://www.cerfacs/algor/Softs.

\bibitem[Freitag(2007)]{Freitag_PhD_07}
{\sc Freitag, M.~A.} 2007 Inner-outer iterative methods for eigenvalue problems
  - convergence and preconditioning. PhD thesis, University of Bath, Bath, U.K.

\bibitem[Golubitsky {\em et~al.\/}({2000})Golubitsky, LeBlanc \&
  Melbourne]{GoLeMe_JNS_00}
{\sc Golubitsky, M., LeBlanc, V.G. \& Melbourne, I.} {2000} {Hopf bifurcation
  from rotating waves and patterns in physical space}. {\em J.\,Nonlinear
  Sci.\/} {\bf {10}}~({1}), {69--101}.

\bibitem[Grossmann(2000)]{Gro00}
{\sc Grossmann, S.} 2000 The onset of shear flow turbulence. {\em Rev.\ Modern
  Phys.\/} {\bf 72}~(2), 603--618.

\bibitem[Guckenheimer \& Holmes(1983)]{Guckenheimer_B_83}
{\sc Guckenheimer, J. \& Holmes, P.} 1983 {\em Nonlinear Oscillations,
  Dynamical Systems, and Bifurcations of Vector Fields\/}. New York:
  Springer-Verlag.

\bibitem[Hof {\em et~al.\/}(2004)Hof, van Doorne, Westerweel, Nieuwstadt,
  Faisst, Eckhardt, Wedin, Kerswell \& Waleffe]{HVWNFEWKW_SCI_04}
{\sc Hof, B., van Doorne, C.W.H., Westerweel, J., Nieuwstadt, F.T.M., Faisst,
  H., Eckhardt, B., Wedin, H., Kerswell, R.R. \& Waleffe, F.} 2004 Experimental
  observation of nonlinear travelling waves in turbulent pipe flow. {\em
  Science\/} {\bf 305}~({5690}), 1594--1598.

\bibitem[Hof {\em et~al.\/}(2006)Hof, Schneider, Westerweel \&
  Eckhardt]{HSWE_NATURE_06}
{\sc Hof, B., Schneider, T.M., Westerweel, J. \& Eckhardt, B.} 2006 Finite
  lifetime of turbulence in shear flows. {\em NATURE\/} {\bf 443}~(7107),
  59--62.

\bibitem[Ioos \& Adelmeyer(1998)]{IoosAdelmeyer_B_98}
{\sc Ioos, G. \& Adelmeyer, M.} 1998 {\em Topics in Bifurcation Theory and
  Applications\/}, 2nd edn. London: World Scientific Publishing.

\bibitem[Kerswell \& Tutty(2007)]{KeTu_JFM_07}
{\sc Kerswell, R.R. \& Tutty, O.R.} 2007 Recurrence of travelling waves in
  transitional pipe flow. {\em J.\,Fluid Mech.\/} {\bf 584}, 69--102.

\bibitem[Knobloch({1986})]{Knobloch_PLA_86}
{\sc Knobloch, E.} {1986} {Normal forms for bifurcations at a double zero
  eigenvalue}. {\em Phys.\ Lett.\,A\/} {\bf {115}}~({5}), {199--201}.

\bibitem[Krupa({1990})]{Krupa_SJMA_90}
{\sc Krupa, M.} {1990} {Bifurcations of relative equilibria}. {\em SIAM
  J.\,Math.\ Anal.\/} {\bf {21}}~({6}), {1453--1486}.

\bibitem[Kuznetsov(1995)]{Kuznetsov_B_95}
{\sc Kuznetsov, Y.A.} 1995 {\em Elements of Applied Bifurcation Theory\/}, 3rd
  edn. New York: Springer-Verlag.

\bibitem[Lehoucq \& Scott(1996)]{ARPACK_96}
{\sc Lehoucq, R. \& Scott, J.~A.} 1996 An evaluation of software for computing
  eigenvalues of sparse nonsymmetric matrices. Technical Report MCS-P547-1195.
  Argonne National Laboratory, http://www.caam.rice.edu/software/ARPACK.

\bibitem[Mamun \& Tuckerman(1995)]{MaTu_PoF_95}
{\sc Mamun, C.K. \& Tuckerman, L.S.} 1995 Asymmetry and {Hopf} bifurcation in
  spherical {Couette} flow. {\em Phys.\ Fluids\/} {\bf 7}~({1}), 80--91.

\bibitem[Mellibovsky \& Meseguer(2009)]{MeMe_PTRSA_09}
{\sc Mellibovsky, F. \& Meseguer, A.} 2009 Critical threshold in pipe flow
  transition. {\em Phil.\ Trans.\ Roy.\ Soc.\ Lond.\,A\/} {\bf 367}~(1888),
  545--560.

\bibitem[Meseguer {\em et~al.\/}(2007)Meseguer, Avila, Mellibovsky \&
  Marques]{MAMM_EPJST_07}
{\sc Meseguer, A., Avila, M., Mellibovsky, F. \& Marques, P.} 2007 Solenoidal
  spectral formulations for the computation of secondary flows in cylindrical
  and annular geometries. {\em Eur.\ Phys.\ J.\, Special Topics\/} {\bf 146},
  249--259.

\bibitem[Meseguer \& Mellibovsky(2007)]{MeMe_ANM_07}
{\sc Meseguer, A. \& Mellibovsky, F.} 2007 On a solenoidal {Fourier-Chebyshev}
  spectral method for stability analysis of the {Hagen-Poiseuille} flow. {\em
  Appl.\ Num.\ Math.\/} {\bf 57}, 920--938.

\bibitem[Meseguer \& Trefethen(2003)]{MeTr_JCP_03}
{\sc Meseguer, A. \& Trefethen, L.N.} 2003 Linearized pipe flow to {Reynolds}
  number $10^7$. {\em J.\,Comput.\ Phys.\/} {\bf 186}, 178--197.

\bibitem[Nagata({1997})]{Nagata_PRE_97}
{\sc Nagata, M.} {1997} {Three-dimensional traveling-wave solutions in plane
  Couette flow}. {\em Phys.\ Rev.\,E\/} {\bf {55}}~({2}), {2023--2025}.

\bibitem[Pfenniger(1961)]{Pfenniger_B_61}
{\sc Pfenniger, W.} 1961 {\em Boundary Layer and Flow Control\/}, chap.
  Transition in the inlet length of tubes at high {Reynolds} numbers, pp.
  970--980. Pergamon.

\bibitem[Pringle {\em et~al.\/}({2009})Pringle, Duguet \&
  Kerswell]{PrDuKe_PTRSA_09}
{\sc Pringle, C.C.T., Duguet, Y. \& Kerswell, R.R.} {2009} Highly symmetric
  travelling waves in pipe flow. {\em Phil.\ Trans.\ Roy.\ Soc.\ Lond.\,A\/}
  {\bf {367}}~({1888}), {457--472}.

\bibitem[Pringle \& Kerswell(2007)]{PrKe_PRL_07}
{\sc Pringle, C.C.T. \& Kerswell, R.R.} 2007 Asymmetric, helical, and
  mirror-symmetric traveling waves in pipe flow. {\em Phys.\ Rev.\ Lett.\/}
  {\bf 99}~(7), 074502.

\bibitem[Pugh \& Saffman({1988})]{PuSa_JFM_88}
{\sc Pugh, J.D. \& Saffman, P.G.} {1988} {Two-dimensional superharmonic
  stability of finite-amplitude waves in plane poiseuille flow}. {\em J.\,Fluid
  Mech.\/} {\bf {194}}, {295--307}.

\bibitem[Quarternoni {\em et~al.\/}(2007)Quarternoni, Sacco \&
  Saleri]{QuSaSa_B_07}
{\sc Quarternoni, A., Sacco, R. \& Saleri, F.} 2007 {\em Numerical
  Mathematics\/}, 2nd edn. Berlin: Springer-Verlag.

\bibitem[Rand({1982})]{Rand_ARMA_82}
{\sc Rand, D.} {1982} {Dynamics and symmetry - Predictions for modulated waves
  in rotating fluids}. {\em Arch.\ Ration.\ Mech.\ An.\/} {\bf {79}}~({1}),
  {1--37}.

\bibitem[Reynolds(1883)]{Rey1883}
{\sc Reynolds, O.} 1883 An experimental investigation of the circumstances
  which determine whether the motion of water shall be direct or sinuous and of
  the law of resistance in parallel channels. {\em Phil.\ Trans.\ Roy.\ Soc.\
  Lond.\/} {\bf 174}, 935--982.

\bibitem[Sanchez {\em et~al.\/}({2002})Sanchez, Marques \&
  Lopez]{SaMaLo_JCP_02}
{\sc Sanchez, J., Marques, F. \& Lopez, J.M.} {2002} {A continuation and
  bifurcation technique for Navier-Stokes flows}. {\em J.\,Comput.\ Phys.\/}
  {\bf {180}}~({1}), {78--98}.

\bibitem[Schmid \& Henningson(1994)]{ScHe_JFM_94}
{\sc Schmid, P.J. \& Henningson, D.S.} 1994 Optimal energy growth in
  {Hagen-Poiseuille} flow. {\em J.\,Fluid Mech.\/} {\bf 277}, 197--225.

\bibitem[Schneider {\em et~al.\/}({2007})Schneider, Eckhardt \&
  Vollmer]{ScEcVo_PRE_07}
{\sc Schneider, T.M., Eckhardt, B. \& Vollmer, J.} {2007} {Statistical analysis
  of coherent structures in transitional pipe flow}. {\em Phys.\ Rev.\,E\/}
  {\bf {75}}~({6}), 066313.

\bibitem[Schneider {\em et~al.\/}(2007)Schneider, Eckhardt \&
  Yorke]{ScEcYo_PRL_07}
{\sc Schneider, T.M., Eckhardt, B. \& Yorke, J.A.} 2007 Turbulence transition
  and edge of chaos in pipe flow. {\em Phys.\ Rev.\ Lett.\/} {\bf 99}~({3}),
  034502.

\bibitem[Shan {\em et~al.\/}(1999)Shan, Ma, Zhang \& Nieuwstadt]{SMZN_JFM_99}
{\sc Shan, H., Ma, B., Zhang, Z. \& Nieuwstadt, F.T.M.} 1999 {On the spatial
  evolution of a wall-imposed periodic disturbance in pipe Poiseuille flow at
  $\rm Re=3000$. Part 1. Subcritical disturbance}. {\em J.\,Fluid Mech.\/} {\bf
  398}, 181--224.

\bibitem[Skufca {\em et~al.\/}(2006)Skufca, Yorke \& Eckhardt]{SkYoEc_PRL_07}
{\sc Skufca, J.D., Yorke, J.A. \& Eckhardt, B.} 2006 Edge of chaos in a
  parallel shear flow. {\em Phys.\ Rev.\ Lett.\/} {\bf 96}~({17}), 174101.

\bibitem[Smale({1967})]{Smale_BAMS_67}
{\sc Smale, S.} {1967} {Differentiable dynamical systems .I. Diffeomorphisms}.
  {\em {Bull.\ Amer.\ Math.\ Soc.}\/} {\bf {73}}~({6}), {747--817}.

\bibitem[Soibelman \& Meiron({1991})]{SoMe_JFM_91}
{\sc Soibelman, I. \& Meiron, D.I.} {1991} {Finite-amplitude bifurcations in
  plane Poiseuille flow - 2-dimensional Hopf-bifurcation}. {\em J.\,Fluid
  Mech.\/} {\bf {229}}, {389--416}.

\bibitem[Takens({1974})]{Takens_IHES_74}
{\sc Takens, F.} {1974} {Singularities of vector fields}. {\em {Publ.\ Math.\
  IHES}\/} {\bf {43}}~({47--100}).

\bibitem[Waleffe({1995})]{Waleffe_SAM_95}
{\sc Waleffe, F.} {1995} {Hydrodynamic stability and turbulence - Beyond
  transients to a self-sustaining process}. {\em Stud.\ Appl.\ Math.\/} {\bf
  {95}}~({3}), {319--343}.

\bibitem[Waleffe(1997)]{Wal_PoF_97}
{\sc Waleffe, F.} 1997 On a self-sustaining process in shear flows. {\em Phys.\
  Fluids\/} {\bf 9}~(4), 883--900.

\bibitem[Wang {\em et~al.\/}({2007})Wang, Gibson \& Waleffe]{WaGiWa_PRL_07}
{\sc Wang, J., Gibson, J. \& Waleffe, F.} {2007} {Lower branch coherent states
  in shear flows: Transition and control}. {\em Phys.\ Rev.\ Lett.\/} {\bf
  {98}}~({20}), 204501.

\bibitem[Wedin \& Kerswell(2004)]{WeKe_JFM_04}
{\sc Wedin, H. \& Kerswell, R.R.} 2004 Exact coherent structures in pipe flow:
  travelling wave solutions. {\em J.\,Fluid Mech.\/} {\bf 508}, 333--371.

\bibitem[Wiggins(2003)]{Wiggins_B_03}
{\sc Wiggins, S.} 2003 {\em Introduction to Applied Nonlinear Dynamical Systems
  and Chaos\/}, 2nd edn. New York: Springer-Verlag.

\bibitem[Willis \& Kerswell({2008})]{WiKe_PRL_08}
{\sc Willis, A.P. \& Kerswell, R.R.} {2008} {Coherent structures in localized
  and global pipe turbulence}. {\em Phys.\ Rev.\ Lett.\/} {\bf {100}}~({12}).

\bibitem[Wygnanski \& Champagne(1973)]{WyCh_JFM_73}
{\sc Wygnanski, I.J. \& Champagne, F.H.} 1973 {On transition in a pipe. Part 1.
  The origin of puffs and slugs and the flow in a turbulent slug}. {\em
  J.\,Fluid Mech.\/} {\bf 59}, 281--335.

\bibitem[Wygnanski {\em et~al.\/}(1975)Wygnanski, Sokolov \&
  Friedman]{WySoFr_JFM_75}
{\sc Wygnanski, I.J., Sokolov, M. \& Friedman, D.} 1975 {On transition in a
  pipe. Part 2. The equilibrium puff}. {\em J.\,Fluid Mech.\/} {\bf 69},
  283--304.

\bibitem[Zikanov(1996)]{Zik_PoF_96}
{\sc Zikanov, O.Y.} 1996 On the instability of pipe {Poiseuille} flow. {\em
  Phys.\ Fluids\/} {\bf 8}~(11), 2923--2932.

\end{thebibliography}

\end{document}